\documentclass[12pt,english,preprint]{revtex4-1}
\usepackage{times}
\usepackage{array}
\usepackage{longtable}
\usepackage{float}
\usepackage{amsmath}
\usepackage{bm}
\usepackage{graphicx}
\usepackage{amssymb}
\usepackage{color}
\usepackage[FIGTOPCAP]{subfigure}

\makeatletter
\usepackage{fancyhdr}
\usepackage{babel}
\makeatother
\begin{document}

\title{A Physics-Constrained Deep Learning Treatment of Runaway Electron Dynamics}

\author{Christopher J. McDevitt}
\email{cmcdevitt@ufl.edu}
\affiliation{Nuclear Engineering Program, Department of Materials Science and Engineering, University of Florida, Gainesville, FL 32611, United States of America}
\author{Jonathan Arnaud}
\affiliation{Nuclear Engineering Program, Department of Materials Science and Engineering, University of Florida, Gainesville, FL 32611, United States of America}
\author{Xian-Zhu Tang}
\affiliation{Theoretical Division, Los Alamos National Laboratory, Los Alamos, NM 87545, United States of America}

\date{\today}

\begin{abstract}

An adjoint formulation leveraging a physics-informed neural network (PINN) is employed to advance the density moment of a runaway electron (RE) distribution forward in time. A distinguishing feature of this approach is that once the adjoint problem is solved, its solution can be used to project the RE density forward in time for an arbitrary initial momentum space distribution of REs. Furthermore, by employing a PINN, a \emph{parametric} solution to the adjoint problem can be learned. Thus, once trained, this adjoint-deep learning framework is able to efficiently project the RE density forward in time across various plasma conditions while still including a fully kinetic description of RE dynamics. As an example application, the temporal evolution of the density of primary electrons is studied, with particular emphasis on evaluating the decay of a RE population when below threshold. Predictions from the adjoint-deep learning framework are found to be in good agreement with a traditional relativistic electron Fokker-Planck solver, for several distinct initial conditions, and across an array of physics parameters. Once trained the PINN thus provides a means of generating RE density time histories with exceptionally low online execution time.


\end{abstract}

\maketitle


\section{Introduction}

The treatment of runaway electron (RE) dynamics is a critical component in the integrated description of tokamak disruptions~\cite{Hender:2007, breizman2019physics} and startup scenarios~\cite{de2023cross, Knoepfel:1979}. While first-principle kinetic solvers are available to evaluate the transient RE dynamics during these phases of a tokamak discharge~\cite{Harvey:2000, Nilsson:2015, mcdevitt2019avalanche, hoppe2021dream, beidler2024wall}, such approaches are computationally intensive, thus sharply limiting their application. This has led to the development of several reduced models of RE formation that may be conveniently coupled to magnetohydrodynamic (MHD) codes~\cite{bandaru2021magnetohydrodynamic, liu2021self, sainterme2024resistive}. These reduced models, however, are often based on the assumption of a slowly evolving background plasma and are thus unable to accurately capture highly transient phases of RE generation or decay. An exception to this trend includes reduced models of hot tail RE seed formation~\cite{smith2008hot, McDevitt:hottail:2023, yang2024pseudoreversible}, where RE formation during a rapid thermal quench is estimated.
Such models, however, have to date been developed specifically for the thermal quench, and thus are not applicable to other transient phases of a disruption, and do not self-consistently include the presence of large-angle collisions. The amplification of the RE seed by large-angle collisions, the so-called RE avalanche~\cite{Sokolov:1979}, is instead incorporated by a separate module~\cite{Rosenbluth:1997, hesslow2019influence, arnaud2024physics} that does not directly account for the transiently evolving momentum space distribution of REs. 

The present work, together with a companion paper~\cite{mcdevittpart22024}, aim to address these limitations by employing an adjoint-deep learning framework that will allow the transient dynamics of REs to be evolved with an online execution time that is orders of magnitude less than a traditional relativistic Fokker-Planck solver. To achieve this, we will first focus on the solution to an adjoint problem~\cite{karney1986current, Liu:2017, zhang2017backward, McDevitt:hottail:2023}, rather than directly solving the relativistic Fokker-Planck equation.
A distinguishing feature of this approach is that a single solution of the adjoint problem can be used to evolve the RE density forward in time for an arbitrary initial momentum space distribution. \emph{Thus, once a solution to the adjoint problem has been identified, the impact of distinct initial momentum space distributions on RE density evolution can be rapidly evaluated for a given set of physics parameters.} The second component of the framework will be the use of a physics-informed neural network (PINN)~\cite{raissi2019physics} to identify a \emph{parametric} solution to the adjoint of the relativistic Fokker-Planck equation. Physics-informed neural networks are a form of deep learning, whereby physical constraints are embedded in the training of a neural network (NN). While this approach has gained prominence due to its ability to integrate physical constraints with data~\cite{karniadakis2021physics}, or treat inverse problems~\cite{cai2021physics}, for cases where a closed system of equations is available, \emph{PINNs can be used to learn the solution to a PDE across a broad range of physics parameters in the absence of data}~\cite{sun2020surrogate, mcdevitt2024knudsen}. Further noting that while the offline training time of a PINN is long, its online inference time is orders of magnitude faster than a traditional solver. The PINN will thus provide a rapid surrogate of the physical system of interest. Combining the above two components, the resulting adjoint-deep learning framework will be able to generate time histories of the RE number density at a fraction of the cost of a traditional solver, for an arbitrary initial momentum space distribution of REs and across a range of plasma conditions. The present paper, along with its companion~\cite{mcdevittpart22024}, will focus on developing the fundamental components of this adjoint-deep learning framework, while future works will focus on employing it for describing tokamak disruptions.

As specific example problems, we consider the decay of a primary RE distribution, while a companion paper~\cite{mcdevittpart22024} will treat the avalanche amplification of an initial RE seed. While large-angle collisions are essential to providing a description of RE generation during the current quench phase of a disruption, an accurate treatment of RE decay is essential when describing the decay of plasma current during the RE plateau. This decay rate will be inferred as a function of electric field strength, effective charge, and the strength of synchrotron radiation. The RE decay rate is shown to have a strong nonlinear dependence on the strength of the electric field, in contrast to commonly employed analytic expressions for RE growth and decay such as Ref. \cite{Rosenbluth:1997}, which indicate a linear dependence. Furthermore, we find that characterizing the decay of REs using a simple decay rate, as is often done in current applications, is often inaccurate either due to the long time required for the RE distribution to settle into a well defined exponential decay, or due to the RE density exhibiting temporal dynamics that cannot be fit by an exponential, even in the long time limit. While the present paper does not include large-angle collisions, the companion paper~\cite{mcdevittpart22024} finds that the decay of the overall RE distribution is well approximated by the decay of the primary population neglecting large-angle collisions.
The present estimate thus provides an efficient means of evaluating the decay of a RE population when below threshold. As a further application, time histories of the RE density evolution are computed for distinct initial momentum space distributions of REs. It is shown that the adjoint-deep learning framework is able to accurately resolve these time histories for RE distributions with sharply different initial pitch distributions. The specific time histories predicted by the adjoint-deep learning framework are compared with those predicted by the particle-based RE solver RAMc~\cite{mcdevitt2019avalanche}, with good agreement evident across a range of initial momentum space distributions.

The remainder of this paper is organized as follows. Section \ref{sec:AFR} discusses the adjoint formulation of the relativistic Fokker-Planck equation. A brief description of PINNs is provided in Sec. \ref{sec:PINN}, along with a discussion of the tailored NN architecture employed in this paper. The temporal evolution of the number density of primary electrons for distinct initial momentum space distributions of REs is given in Sec. \ref{sec:DPR}, along with evaluating RE decay for a range of plasma parameters. A brief discussion along with conclusions are given in Sec. \ref{sec:C}.

\section{\label{sec:AFR}Adjoint Framework for Relativistic Electron Evolution}


\subsection{\label{sec:LDCF}Adjoint of the Relativistic Fokker-Planck Equation}

The adjoint to the relativistic Fokker-Planck equation has received substantial treatment throughout the plasma physics literature, both in the context of wave driven currents~\cite{antonsen1982radio, taguchi1983effect, fisch1986transport, karney1986current} along with more recent treatments focusing on RE formation~\cite{Liu:2016, Liu:2017, zhang2017backward, McDevitt:hottail:2023, arnaud2024physics}. In this section we will describe an adjoint problem,
where our interest will be evaluating the probability that an electron initially located at a given energy and pitch, runs away at a later time. 
We begin by introducing the Green's function $F \left( \mathbf{p}, t;\mathbf{p}_0, t_0 \right)$, which is taken to obey the relativistic Fokker-Planck equation:
\begin{subequations}
\label{eq:TDRP4}
\begin{align}
\frac{\partial F \left( \mathbf{p}, t;\mathbf{p}_0, t_0 \right)}{\partial t} + E \left( F, t\right) + C \left( F, t\right) + R \left( F, t\right) = \delta \left( \mathbf{p} - \mathbf{p}_0 \right) \delta \left( t - t_0 \right)
, \label{eq:TDRP4a}
\end{align}
where $F$ evolves due to electric field acceleration, small-angle collisions, and losses from synchrotron radiation, 
whose form are defined by:
\begin{equation}
E \left( F, t\right) = -\frac{1}{p^2} \frac{\partial}{\partial p} \left[ p^2 E_\Vert \xi F \left( \mathbf{p}, t;\mathbf{p}_0, t_0 \right) \right] - \frac{\partial}{\partial \xi} \left[ \left( \frac{1-\xi^2}{p}\right) E_\Vert F \left( \mathbf{p}, t;\mathbf{p}_0, t_0 \right) \right]
, \label{eq:TDRP4b}
\end{equation}
\begin{align}
C \left( F, t\right) &= -\frac{1}{p^2} \frac{\partial}{\partial p} p^2  \left[ C_F F \left( \mathbf{p}, t;\mathbf{p}_0, t_0 \right) \right] - \frac{C_B}{p^2} \frac{\partial}{\partial \xi} \left[ \left( 1-\xi^2\right) \frac{\partial}{\partial \xi} F \left( \mathbf{p}, t;\mathbf{p}_0, t_0 \right) \right]
, \label{eq:TDRP4c}
\end{align}
\begin{equation}
R \left( F, t\right) = - \frac{1}{p^2} \frac{\partial}{\partial p} \left[ \alpha p^3 \gamma \left( 1 - \xi^2 \right) F \left( \mathbf{p}, t;\mathbf{p}_0, t_0 \right) \right] + \frac{\partial}{\partial \xi} \left[ \alpha \frac{\xi \left( 1-\xi^2 \right)}{\gamma} F \left( \mathbf{p}, t;\mathbf{p}_0, t_0 \right) \right]
. \label{eq:TDRP4d}
\end{equation}
\end{subequations}
Here, $\left( \mathbf{p}_0, t_0\right)$ indicate the momentum and time the electron was injected at, we will consider a bounded system with momentum in the range $p\in \left[ p_{min}, p_{max}\right]$, the electron's pitch is defined by $\xi \equiv p_\Vert / p \in \left[ -1, 1 \right]$, time is normalized as $t \to t/\tau_c$, with $\tau_c \equiv 4\pi \epsilon^2_0 m^2_e c^3 / \left( e^4 n_e \ln \Lambda \right)$, $\ln \Lambda$ is the Coulomb logarithm, momentum as $p\to p / \left( m_e c \right)$, the parallel electric field as $E_\Vert \to E_\Vert / E_c$, where $E_c \equiv m_e c / \left( e\tau_c\right)$ is the Connor-Hastie electric field~\cite{Connor:1975}, the Lorentz factor is defined as $\gamma = \sqrt{1+p^2}$, $\alpha \equiv \tau_c/\tau_s$, and $\tau_s \equiv 6\pi \epsilon_0 m^3_e c^3 / \left( e^4 B^2\right)$ is the timescale associated with synchrotron radiation. The coefficients for the collisional drag $C_F$ and pitch-angle scattering $C_B$ in Eq. (\ref{eq:TDRP4c}) are defined by:
\begin{subequations}
\begin{equation}
C_F \equiv \frac{\gamma^2}{p^2}
, \label{eq:TDRP5a}
\end{equation}
\begin{equation}
C_B \equiv \frac{\gamma}{2} \frac{  \left( Z_{eff} + 1 \right)}{p}
. \label{eq:TDRP5b}
\end{equation}
\end{subequations}
We have neglected energy diffusion since it has a magnitude of order $T_e / \left( m_e c^2\right)$ at relativistic energies, such that it will be subdominant for post thermal quench plasmas with temperatures of tens of electron volts or less~\cite{Hender:2007}.

Turning to the boundary conditions on $F \left( \mathbf{p}, t;\mathbf{p}_0, t_0 \right)$, from Eq. (\ref{eq:TDRP4}) the momentum space flow $U_p$ of electrons through the upper and lower momentum boundaries is given by:
\begin{equation}
U_p \equiv -E_\Vert \xi - C_F - \alpha p \gamma \left( 1 - \xi^2 \right)
. \label{eq:TDRP8}
\end{equation}
When $E_\Vert > 1$, regions with both $U_p>0$ and $U_p<0$ will be present at $p=p_{max}$. To ensure that no inflow of electrons through the upper momentum boundary is present we will require $F$ to vanish at $p=p_{max}$ when $U_p < 0$. With regard to the lower momentum boundary, we will ensure that $p_{min}$ is chosen to be small enough such that $U_p<0$ at this surface (leading to a flow directed out of the domain), so no boundary condition on $F$ needs to be applied at $p=p_{min}$. With these boundary conditions, a particle balance relation can be derived by integrating Eq. (\ref{eq:TDRP4a}) over momentum and time, yielding
\begin{align}
\int d^3 p F \left( \mathbf{p}, t_{final};\mathbf{p}_0, t_0 \right) + 2\pi \int^{t_{final}}_{-\infty} dt \int\limits_{\hat{\mathbf{n}}\cdot \mathbf{U}_p > 0} d \xi \left. \left[ p^2 U_p F \left( \mathbf{p}, t;\mathbf{p}_0, t_0 \right) \right] \right|^{p_{max}}_{p_{min}} = 1
, \label{eq:TDRP7}
\end{align}
where the momentum integration is for $p_{min} \leq p \leq p_{max}$, $\hat{\mathbf{n}}$ is a unit vector pointing out of the bounded domain, $\mathbf{U}_p = \hat{\mathbf{p}} U_p$, and we only include regions on the upper momentum boundary with an outward flux of particles such that $\hat{\mathbf{n}}\cdot \mathbf{U}_p > 0$, since $F$ will vanish otherwise.
Noting that $F$ describes the electron distribution resulting from a unit source at $\left( \mathbf{p}_0, t_0 \right)$, Eq. (\ref{eq:TDRP8}) enforces that the probability of the electron remaining inside the momentum space domain bounded by $p_{min}$ and $p_{max}$ [first term in Eq. (\ref{eq:TDRP7})], or exiting through either the low or high momentum boundaries [second term in Eq. (\ref{eq:TDRP7})], is one.

Considering now the adjoint to the relativistic Fokker-Planck equation~\cite{karney1986current, Liu:2017, zhang2017backward, McDevitt:hottail:2023}, which may be expressed as:
\begin{subequations}
\label{eq:TDRP9}
\begin{align}
\frac{\partial P}{\partial t} - E^* \left( P, t\right) - C^* \left( P, t\right) - R^* \left( P, t\right) = 0
, \label{eq:TDRP9a}
\end{align}
with the adjoint operators defined by:
\begin{equation}
E^* \left( P, t\right) = E_\Vert \xi \frac{\partial P}{\partial p} + \left( \frac{1-\xi^2}{p}\right) E_\Vert \frac{\partial P}{\partial \xi}
, \label{eq:TDRP9b}
\end{equation}
\begin{equation}
C^* \left( P, t\right) = C_F \frac{\partial P}{\partial p} - \frac{C_B}{p^2} \frac{\partial}{\partial \xi} \left[ \left( 1-\xi^2\right) \frac{\partial P}{\partial \xi} \right]
, \label{eq:TDRP9c}
\end{equation}
\begin{equation}
R^* \left( P, t\right) = \alpha p \gamma \left( 1 - \xi^2 \right) \frac{\partial P}{\partial p} - \alpha \frac{\xi \left( 1-\xi^2 \right)}{\gamma} \frac{\partial P}{\partial \xi}
. \label{eq:TDRP9d}
\end{equation}
\end{subequations}
By successive integration by parts, an adjoint relation can be written as:
\begin{align}
\int^{t_{final}}_{-\infty} dt \int d^3 p P &\left[ E \left( F, t \right) + C \left( F, t\right) + R \left( F, t\right) \right] \nonumber \\
& = \int^{t_{final}}_{-\infty} dt \int d^3 p F \left[ E^* \left( P, t \right) + C^* \left( P, t\right) + R^* \left( P, t\right) \right] \nonumber \\
& + 2 \pi \int^{t_{final}}_{-\infty} dt \int\limits_{\hat{\mathbf{n}}\cdot \mathbf{U}_p > 0} d \xi \left. \left[ p^2 P U_p F \right] \right|^{p_{max}}_{p_{min}}
, \label{eq:TDRP12}
\end{align}
where we have noted that the pitch flux vanishes at $\xi = \pm 1$, leaving only the fluxes at the upper and lower momentum boundaries [last term in Eq. (\ref{eq:TDRP12})]. Inserting Eqs. (\ref{eq:TDRP4a}) and (\ref{eq:TDRP9a}) into Eq. (\ref{eq:TDRP12}), yields
\begin{align}
\int^{t_{final}}_{-\infty} dt \int d^3 p P &\left[ \delta \left( \mathbf{p} - \mathbf{p}_0 \right) \delta \left( t - t_0 \right) - \frac{\partial F}{\partial t} \right] \nonumber \\
& = \int^{t_{final}}_{-\infty} dt \int d^3 p F \frac{\partial P}{\partial t} + 2\pi \int^{t_{final}}_{-\infty} dt \int\limits_{\hat{\mathbf{n}}\cdot \mathbf{U}_p > 0} d \xi \left. \left[ p^2 P U_p F \right] \right|^{p_{max}}_{p_{min}}
. \label{eq:TDRP13}
\end{align}
Evaluating the delta functions and combining the time derivative terms, yields an expression for $P$, i.e.
\begin{align}
P \left( \mathbf{p}=\mathbf{p}_0, t=t_0 ; t_{final} \right) &= \int d^3 p P \left( t=t_{final} \right) F \left( t=t_{final}\right) \nonumber \\
& + 2\pi \int^{t_{final}}_{-\infty} dt \int\limits_{\hat{\mathbf{n}}\cdot \mathbf{U}_p > 0} d \xi \left. \left[ p^2 P U_p F \right] \right|^{p_{max}}_{p_{min}}
. \label{eq:TDRP14}
\end{align}
The physical meaning of the quantity $P$ is determined by the choice of the terminal condition $P \left( t=t_{final} \right)$, and the upper and lower momentum boundary conditions on $P$. 
While a range of quantities of interest could be evaluated using this adjoint formulation by considering different terminal and momentum space boundary conditions, for the remainder of this paper we will focus on the runaway probability function (RPF)~\cite{karney1986current}. 
This is accomplished
by enforcing the terminal and momentum space boundary conditions:
\begin{subequations}
\label{eq:RPF1}
\begin{equation}
P \left( t=t_{final} \right) = \Theta \left( p - p_{RE}\right)
, \label{eq:RPF1a}
\end{equation}
\begin{equation}
P \left( p=p_{min} \right) = 0
, \label{eq:RPF1b}
\end{equation}
\begin{equation}
P \left( p=p_{max} \right) =
\begin{cases}
1, & U_p \left( p=p_{max} \right) > 0 \\
\text{unconstrained}, & U_p \left( p=p_{max} \right) < 0 
\end{cases}
, \label{eq:RPF1c}
\end{equation}
\end{subequations}
where $\Theta \left( x\right)$ is a Heaviside function and $U_p$ is defined by Eq. (\ref{eq:TDRP8}) above.
Here, $p_{RE}$ is taken to be the momentum an electron must exceed to be labelled as a RE. While there is some degree of arbitrariness in its precise choice, it should be chosen to be larger than the momentum of a thermal electron, but not too much larger than the critical energy to run away to avoid undercounting REs. The lower momentum space boundary condition Eq. (\ref{eq:RPF1b}) is chosen such that electrons that fall through the low momentum boundary, and are thus lost to the bulk plasma, are not counted as REs. The upper momentum space boundary condition Eq. (\ref{eq:RPF1c}) is selected such that electrons with $U_p \left( p=p_{max}\right)> 0$ (which will be accelerated through the upper momentum boundary), are counted as REs. However, for 
regions on the upper momentum boundary where $U_p \left( p=p_{max} \right) < 0$, we will \emph{not} enforce $P \left( p=p_{max} \right) = 1$ since these electrons would not be immediately accelerated through the upper momentum boundary. Furthermore, we also will not set $P \left( p=p_{max} \right) = 0$ for $U_p \left( p=p_{max} \right) < 0$, since these electrons still have a finite probability of running away at a later time. This portion of the upper momentum boundary will thus be left unconstrained. We note that its contribution to the surface term in Eq. (\ref{eq:TDRP14}) will be zero, since $F$ vanishes at the upper momentum boundary for $U_p<0$. 

With the terminal condition and upper and lower momentum boundary conditions described by Eq. (\ref{eq:RPF1}), Eq. (\ref{eq:TDRP14}) can be written as:
\begin{align}
P \left( \mathbf{p}=\mathbf{p}_0, t=t_0 ; t_{final} \right) & = \int d^3 p \Theta \left( p - p_{RE} \right) F \left( t=t_{final}\right) \nonumber \\
& + 2\pi p^2_{max} \int^{t_{final}}_{-\infty} dt \int\limits_{\xi_{out}} d\xi U_p \left( p=p_{max}  \right) F \left( p=p_{max} \right)
, \label{eq:TDRP15}
\end{align}
where $\xi_{out}$ defines the region in pitch that satisfies
\[
U_p \left( p_{max}\right) = -\xi_{out} E_\Vert - C_F \left( p=p_{max} \right) - \alpha p_{max} \gamma_{max} \left( 1-\xi^2_{out} \right) > 0
.
\] 
Comparing Eq. (\ref{eq:TDRP15}) with Eq. (\ref{eq:TDRP7}) implies that $P \left( \mathbf{p}=\mathbf{p}_0, t=t_0; t_{final} \right)$ is the probability that an electron either remains in the simulation domain, but with an energy greater than $p_{RE}$ [first term in Eq. (\ref{eq:TDRP15})], or that the electron has exited through the upper momentum boundary [second term in Eq. (\ref{eq:TDRP15})] by a time $t_{final}$. The quantity $P \left( \mathbf{p}=\mathbf{p}_0, t=t_0 \right)$ thus represents the probability of an electron injected at $\mathbf{p}=\mathbf{p}_0$ and $t=t_0$ obtaining a momentum of $p_{RE}$ or greater at $t_{final}$ and is often referred to as a runaway probability function~\cite{karney1986current, Liu:2016, Liu:2017, zhang2017backward, McDevitt:hottail:2023}. 


\begin{figure}
\begin{centering}
\subfigure[]{\includegraphics[scale=0.33]{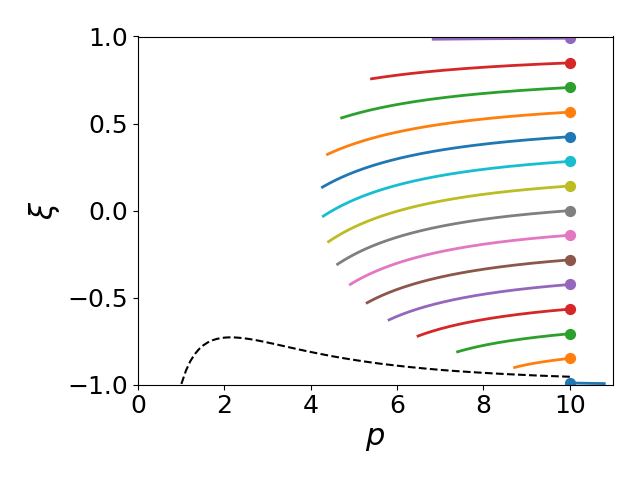}}
\subfigure[]{\includegraphics[scale=0.33]{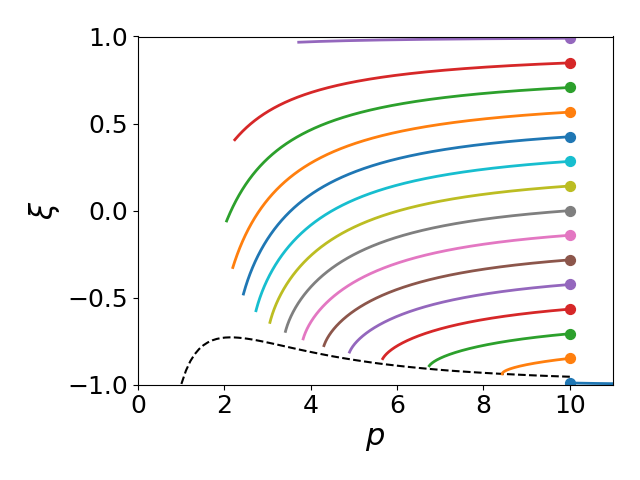}}
\subfigure[]{\includegraphics[scale=0.33]{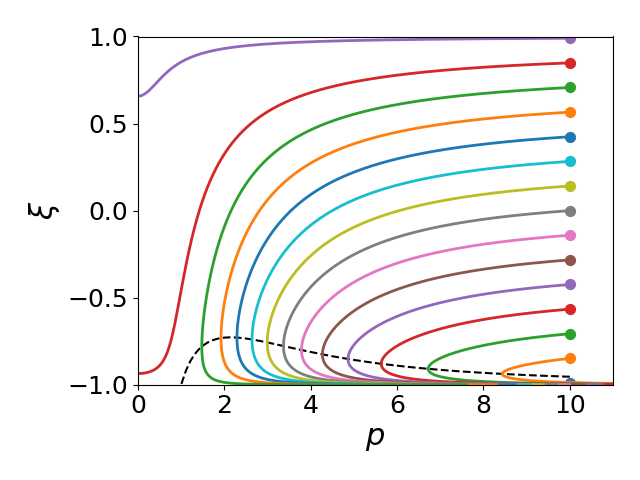}}
\par\end{centering}
\caption{Example deterministic electron orbits. Here, electrons are initialized at the solid circle markers located at $p=10$ and evolved forward in time including electric field acceleration, collisional drag, and synchrotron radiation. Panel (a) indicates the orbits at $t=\tau_c$, panel (b) for $t=2\tau_c$ and panel (c) is for $t=10\tau_c$. The dashed black curve indicates the contour $U_p=0$, where regions below this curve will have $U_p>0$. The parameters were taken to be $E_\Vert = 2$ and $\alpha=0.1$, where $Z_{eff}$ does not enter since pitch-angle scattering was neglected.}
\label{fig:RPF1}
\end{figure}

The physical interpretation of the RPF can be made clear by
considering the orbits of several electrons initialized at different
pitches as shown in Fig. \ref{fig:RPF1}. These orbits have been
computed using characteristic equations derived from
Eq. (\ref{eq:TDRP4}), where pitch-angle scattering was neglected
leading to deterministic orbits. The parameters ($E_\Vert = 2$ and
$\alpha = 0.1$) were selected such that the system is slightly above
the avalanche threshold. After integrating forward in time until
$t=\tau_c$ [see Fig. \ref{fig:RPF1}(a)], all of the electron orbits
have slowed down, except an orbit with an initial pitch of
$\xi=-0.99$. This last orbit was initialized with a pitch that put it
in a region with $U_p>0$ (below the dashed black curve) allowing it to
be accelerated. It is useful to note that the rate at which electrons
in the $U_p<0$ region decelerate depends sensitively on their pitch,
with electrons near $\xi \approx 0$ decelerating the most rapidly due
to synchrotron radiation being maximal in this region. At a later time
[$t=2\tau_c$ Fig. \ref{fig:RPF1}(b)] the majority of electron orbits
continue to slow down, but are also tilted toward the negative pitch
region. This is due to the electric field pushing the electrons toward
the $\xi=-1$ axis. Finally by $t=10\tau_c$, all of the electrons have
either been accelerated through the $p=10$ surface, or decelerated to
the bulk (i.e. $p=0$ since we have assumed a cold bulk plasma). Those
that have been accelerated through the $p=10$ surface [which could be
  taken to be the upper momentum boundary] will be treated as REs
according to Eq. (\ref{eq:TDRP15}).
In contrast, two orbits in Fig. \ref{fig:RPF1}(c), the purple and red orbits initialized with $\xi > 0.8$ have fallen back to the bulk, and are thus deemed to not run away. Thus, for the case where pitch-angle scattering is neglected, a separatrix will form delineating runaway and non-runaway orbits.

\begin{figure}
\begin{centering}
\subfigure[]{\includegraphics[scale=0.33]{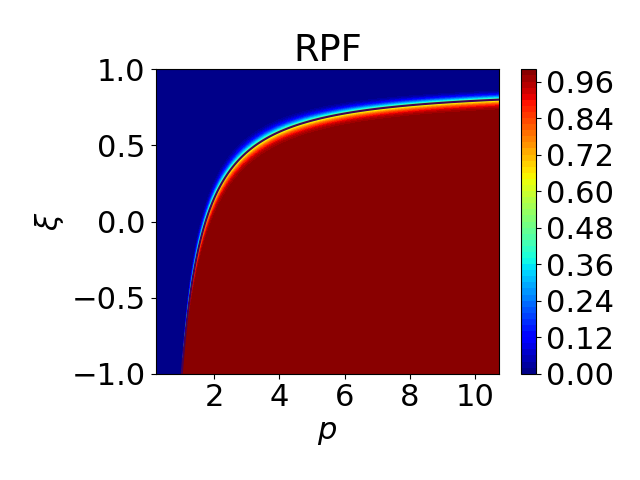}}
\subfigure[]{\includegraphics[scale=0.33]{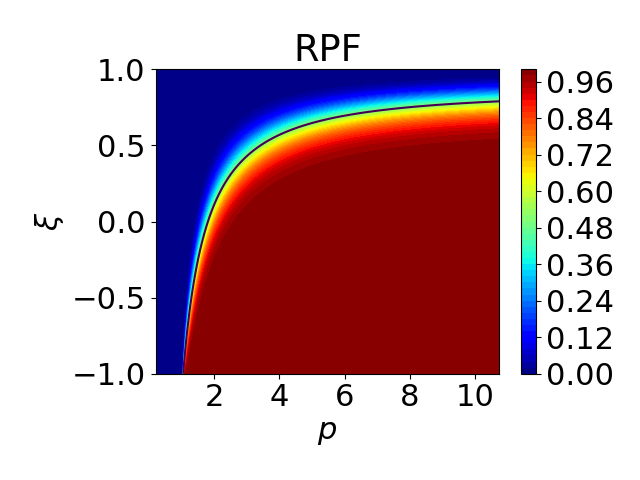}}
\subfigure[]{\includegraphics[scale=0.33]{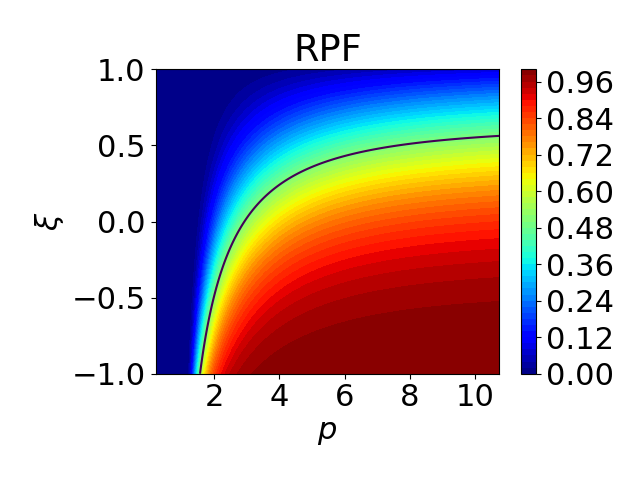}}
\par\end{centering}
\caption{Steady state RPF with the pitch-angle scattering term multiplied by a factor of $0.01$ [panel (a)], $0.1$ [panel (b)], and $1$ [panel (c)]. The parameters were taken to be $E_\Vert = 2$, $Z_{eff}=1$ and $\alpha=0.1$. The RPF was evaluated by solving Eq. (\ref{eq:TDRP9}) at steady state using a physics informed neural network, whose specific implementation is described in Ref. \cite{arnaud2024physics}.}
\label{fig:RPF2}
\end{figure}

The primary impact of pitch-angle scattering will be to broaden the sharp transition region and shift its location to higher energy. This is illustrated in Fig. \ref{fig:RPF2}, where we have computed the RPF from the steady state adjoint equation, but artificially multiplied the pitch-angle scattering term by factors of $0.01$ [Fig. \ref{fig:RPF2}(a)], $0.1$ [Fig. \ref{fig:RPF2}(b)] and $1$ [Fig. \ref{fig:RPF2}(c)] for a plasma with $Z_{eff}=1$, while keeping all other terms fixed, including the collisional drag. It is evident that for Fig. \ref{fig:RPF2}(a) (where the pitch-angle scattering term is multiplied by $0.01$), a sharp transition region is present, where the region with $P\approx 1$ closely aligns with the RE orbits shown in Fig. \ref{fig:RPF1}(c). However, as the strength of pitch-angle scattering is increased in Figs.~\ref{fig:RPF2}(b) and 2(c), the transition region broadens, and the location of the $P=0.5$ contour shifts to higher energy, implying that RE generation will be less efficient as pitch-angle scattering is increased.

\subsection{\label{sec:REDE}Runaway Electron Density Evolution}

An immediate application of the RPF is to predict the number of REs at a later time. Specifically, if an electron distribution $f^{(init)}_e \left( p,\xi \right)$ was present at $t=0$, the number density of electrons at a later time $t_{final}$ is given by:
\begin{equation}
n_{RE} \left( t_{final}\right) = \int d^3 p f^{(init)}_e \left( p, \xi \right) P \left( p, \xi, t=0 ; t_{final} \right)
. \label{eq:TDRP16}
\end{equation}
Here, $P \left( p, \xi, t=0 ; t_{final} \right)$ 
indicates the probability that an electron initially located at $\left( p, \xi \right)$ runs away after a period of duration $t_{final}$. Hence, by weighing the density moment of the initial distribution of electrons by $P \left( p, \xi, t=0 ; t_{final} \right)$ this yields the number density of REs at $t=t_{final}$. 

\begin{figure}
\begin{centering}
\includegraphics[scale=0.5]{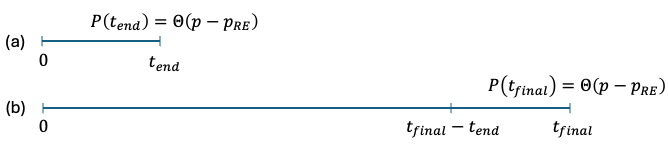}
\par\end{centering}
\caption{Panel (a) indicates an adjoint problem with the terminal condition $P \left( t=t_{end} \right) = \Theta \left( p-p_{RE} \right)$, and Eq. (\ref{eq:TDRP9}) is integrated backward to $t=0$. The solution to this problem will be identical to a problem with the terminal condition at $P \left( t=t_{final} \right) = \Theta \left( p-p_{RE} \right)$ and integrating backward until $t=t_{final}-t_{end}$.}
\label{fig:RED1}
\end{figure}

The number density of REs at times intermediate between $t=0$ and $t=t_{final}$ can be obtained by a generalization of Eq. (\ref{eq:TDRP16}). A conceptually simple generalization would be to solve a series of terminal value problems for end times $t_{end}$, where $t_{end}$ falls between $t=0$ and $t=t_{final}$ [see Fig. \ref{fig:RED1}(a)]. Each of these problems would take $P \left( t=t_{end}\right) = \Theta \left( p-p_{RE}\right)$ as the terminal condition and integrate backward to $t=0$, thus allowing the RE density to be projected to an end time $t_{end}$, such that Eq. (\ref{eq:TDRP16}) would then be written as:
\begin{equation}
n_{RE} \left( t_{end}\right) = \int d^3 p f^{(init)}_e \left( p, \xi \right) P \left( p, \xi, t=0 ; t_{end}\right)
, \label{eq:TDRP18}
\end{equation}
where $f^{(init)}_e \left( p, \xi \right)$ is the electron distribution at $t=0$. While conceptually straightforward, this approach would involve a substantial increase in computational cost, since several terminal value problems must be solved. A more efficient means of determining the RE density at intermediate times between $t=0$ and $t_{final}$ can be arrived at by noting that solving a terminal value problem ending at $t=t_{end}$, and integrating backward until $t=0$, is identical to defining the terminal value at $t_{final}$ and integrating backward for a period of duration $t_{final}-t_{end}$ [see Fig. \ref{fig:RED1}(b)]. In this latter case, the RE density at $t_{end}$ is given by:
\begin{equation}
n_{RE} \left( t_{end} \right) = \int d^3 p f^{(init)}_e \left( p, \xi \right) P \left( p, \xi, t_{final}-t_{end} ; t_{final}\right)
. \label{eq:TDRP19}
\end{equation}
Hence, by solving a terminal value problem with $P \left( t_{final} \right) = \Theta \left( p-p_{RE}\right)$ and integrating backward until $t=0$, the RPF at all intermediate times will be computed. Once this terminal value problem has been solved, the RE density at a time between $0 \leq t \leq t_{final}$ is given by:
\begin{equation}
n_{RE} \left( t \right) = \int d^3 p f^{(init)}_e \left( p, \xi \right) P \left( p, \xi, \tau ; t_{final} \right)
, \label{eq:TDRP20}
\end{equation}
where $\tau \equiv t_{final} - t$. Equation (\ref{eq:TDRP20}) will allow the time evolution of the density moment of an initial seed population of REs with distribution $f^{(init)}_e \left( p, \xi \right)$ to be projected to any time between $0$ and $t_{final}$.

\section{\label{sec:PINN}Physics-informed Neural Networks}

\subsection{\label{sec:PCDL}Physics-Constrained Deep Learning}

The adjoint problem discussed in Sec. \ref{sec:AFR} allows the RE density to be evolved in time beginning with an arbitrary initial momentum space distribution of REs. Our aim in this section will be to extend the generalizability of the adjoint framework by utilizing a PINN to evaluate the parametric solution to Eq. (\ref{eq:TDRP9}). In this way, the resulting adjoint-deep learning framework will be able to \emph{evaluate the RE density for both an arbitrary initial electron distribution, and a range of plasma conditions.} Once trained, the resulting framework will allow for nearly instantaneous projections of RE density incorporating fully kinetic physics.

Physics-informed neural networks~\cite{raissi2019physics, karniadakis2021physics} have emerged as a prominent example of physics-constrained deep learning methods~\cite{lagaris1998artificial, karpatne2017theory, lusch2018deep, wang2020towards}.
The present discussion will only focus on the essential concepts, where the interested reader is referred to Ref. \cite{karniadakis2021physics} (and references therein) for a more detailed discussion. Physics-informed neural networks utilize physical constraints to regularize the training of a NN. 
A PINN in its simplest form can be expressed as~\cite{raissi2019physics}:
\begin{equation}
\text{Loss} = \frac{1}{N_{data}} \sum^{N_{data}}_i \left[ P_i - P \left( \mathbf{p}_i, t_i; \bm{\lambda}_i \right)\right]^2 + \frac{w_{PDE}}{N_{PDE}} \sum^{N_{PDE}}_i  \mathcal{R}^2 \left( \mathbf{p}_i, t_i; \bm{\lambda}_i \right)
, \label{eq:PCDL2}
\end{equation}
where $w_{PDE}$ is a scalar that determines the weight given to the PDE portion of the training, $P_i$ represents data points for the quantity of interest (the RPF in the present paper), $\mathbf{p}_i$ are the momentum space coordinates, $t_i$ is time, and $\bm{\lambda}_i$ represent parameters of the physical system which in the present paper will be taken to be $\bm{\lambda} = \left( E_\Vert, Z_{eff}, \alpha \right)$, $N_{data}$ is the number of data points used in the training of the NN, which includes boundary and initial conditions for the PDE as well as available experimental or simulation data, $N_{PDE}$ is the number of points where the physical constraint is sampled, and $\mathcal{R}\left( \mathbf{p}_i, t; \bm{\lambda}_i \right)$ is the residual of the PDE, which penalizes deviations from the imposed physical constraint of the system. We note that while the amount of experimental or simulation data is often limited, the value of $N_{PDE}$ may be taken to be very large, where increasing its value will only impact the computational cost of the offline training of the model. This property will be exploited to enable the training of the model across a broad range of physical parameters.

For the present study we will consider the limit where no data is available, such that we will rely exclusively on physical constraints for training the PINN. A key component when developing a robust PINN will be to enforce several properties of the solution as hard constraints in the NN architecture itself, rather than as soft constraints in the loss function. This prevents the PINN from making unphysical predictions while simultaneously, and perhaps more importantly, severely restricting the solutions that the optimizer searches across, leading to more robust convergence of the PINN. These hard constraints will be enforced by a customized `physics layer,' detailed in Sec. \ref{sec:PRED} below.

\subsection{\label{sec:PRED}Physics Layer}

As a means of improving the accuracy and robustness of the PINN, we will implement several properties of the solution as hard constraints via the addition of a `physics layer' to the NN. This additional layer will take the output of the hidden layers of the NN and impose a series of transforms that enforce (i) the terminal condition given by Eq. (\ref{eq:RPF1a}), (ii) lower momentum boundary condition [Eq. (\ref{eq:RPF1b})], and (iii) constrains the RPF to have a range between zero and one. While all of these conditions can be enforced by adding terms into the loss function, the use of a physics layer will act to both simplify the form of the loss function, as well as restrict the range of solutions that the optimizer is allowed to search across. The form of the physics layer that we will introduce is given by:
\begin{align}
P^\prime \left( p, \xi, t ; \bm{\lambda}\right) &= P_{term} \left( p \right) + \left( \frac{p-p_{min}}{p_{max}-p_{min}} \right) \tanh \left( t_{final}-t \right) P_{NN} \left( p, \xi, t ; \bm{\lambda} \right)
, \label{eq:PRED1}
\end{align}
where $P_{NN}$ is the output of the hidden layers of the neural network, $\bm{\lambda}$ is a vector containing the physics parameters $\bm{\lambda} = \left( E_\Vert, Z_{eff}, \alpha \right)$, and the terminal condition is defined by
\begin{equation}
P_{term} \left( p \right) \equiv \frac{1}{\Delta P} \tanh \left( \frac{p-p_{RE}}{\Delta p} \right)
, \label{eq:PRED2}
\end{equation}
which transitions from a large negative number to a large positive number as $p$ crosses $p_{RE}$, with $\Delta p$ determining how sharp the transition is, and $\Delta P$ setting the maximum magnitude. The RPF is computed by passing Eq. (\ref{eq:PRED1}) through
\begin{equation}
P \left( p, \xi, t ; \bm{\lambda} \right) = \frac{1}{2} \left\{ 1 +  \tanh \left[ P^\prime \left( p, \xi, t ; \bm{\lambda} \right) \right] \right\}
. \label{eq:PRED6}
\end{equation}
Here, it may be verified that this additional layer to the NN ensures the RPF has a range between zero and one [achieved via Eq. (\ref{eq:PRED6})], approximates the terminal condition of a Heaviside function centered about $p_{RE}$ at $t=t_{final}$ (for $\Delta p \ll 1$ and $\Delta P \ll 1$), and obtains a vanishingly small value at the low momentum boundary when $\Delta p/p_{max} \ll 1$ and $\Delta P \ll 1$. The physics layer contains the hyperparameters $\Delta P$ and $\Delta p/p_{max}$. Each of these parameters should be set to a value much less than one to closely adhere to the physics problem. In particular, $\Delta P$ must be chosen to be sufficiently small such that $P$ has a negligible value at $p=p_{min}$. The parameter 
$\Delta p$ must be small to ensure a sharp interface at $p=p_{RE}$. However, too small a value of $\Delta p$ has a negative impact on training, thus 
reducing the accuracy of the PINN's approximation to the RPF. It's value will thus be chosen by a compromise between adhering to the physics problem where $\Delta p \to 0$, and the accuracy of the PINN, which for too small a value of $\Delta p$ will struggle to resolve the initial transient evolution of the sharp interface (see Fig. \ref{fig:TER2} below). For all the studies shown below we have chosen $\Delta P = 0.15$ and $\Delta p = 0.1 p_{max}$.

With the physics layer defined by Eqs. (\ref{eq:PRED1})-(\ref{eq:PRED6}), the loss function will contain a term proportional to the mean-square-error of the residual of the adjoint to the relativistic Fokker-Planck equation, along with a term penalizing deviations from the upper momentum space boundary condition defined by Eq. (\ref{eq:RPF1c}). The weighting of the different energy and parameter regions will have a strong impact on the performance of the PINN. In this study we will consider a loss function of the form:
\begin{equation}
\text{loss} = \frac{w_{PDE}}{N_{PDE}} \sum^{N_{PDE}}_i \left[ G \left( p \right) \left( \frac{p^2_i}{1+p^2_i} \right) \mathcal{R} \left( p_i,\xi_i,t_i ; \bm{\lambda}_i\right) \right]^2 + \frac{1}{N_{bdy}} \sum^{N_{bdy}}_i \left[ P_i - P \left( p_i,\xi_i,t_i ; \bm{\lambda}_i\right) \right]^2
, \label{eq:PRED7}
\end{equation}
where
\begin{equation}
G \left( p \right) = 1 - \exp \left[ -\frac{\left( p_{max}-p \right)^2}{\Delta p_{max}^2} \right]
, \label{eq:PRED8}
\end{equation}
Here, 
$N_{bdy}$ is the number of points on the upper momentum boundary that will be sampled, and $\mathcal{R}$ is the residual of Eq. (\ref{eq:TDRP9a}). The factor $p^2_i/\left( 1+p^2_i\right)$ in front of the residual is designed to partially cancel the $1/p^3$ divergence in the pitch-angle scattering operator, with the $p-p_{min}$ factor in Eq. (\ref{eq:PRED1}) ensuring the loss doesn't diverge for small $p$. In addition we have multiplied the residual of the PDE by the function $G \left( p\right)$. This function vanishes at $p=p_{max}$ and increases to one for $p<p_{max}$ over a momentum scale defined by $\Delta p_{max}$, where $\Delta p_{max} \ll p_{max}$. The motivation for introducing this factor is that for $E_\Vert$ slightly greater than one, the high momentum boundary condition is $P = 1$ at $\xi=-1$. However, for $E_\Vert$ only slightly greater than one, $P$ will quickly drop to value far below one, particularly in the presence of synchrotron radiation and large $Z_{eff}$. This sharp variation often leads to a large residual localized near $p\approx p_{max}$ and $\xi \approx -1$. Even when localized to a very small region of momentum space, this large residual has a negative impact on training, particularly when residual based adaptive sampling techniques are used. The factor $G \left( p \right)$ will thus reduce the weight placed on the high momentum region, and thus help the PINN obtain good accuracy across the majority of momentum space. For all cases in this paper we will take $\Delta p_{max}=0.05 p_{max}$, such that only the region very close to $p_{max}$ is impacted. Finally, the factor $w_{PDE}$ determines the relative weight of the PDE versus the boundary condition. For all cases we will take $w_{PDE} = 10$, to emphasize the PDE term in the loss function. The Python script used for training the PINN is written using the DeepXDE library \cite{lu2021deepxde} with TensorFlow~\cite{abadi2016tensorflow} as the backend, and will be made available upon acceptance at https://github.com/cmcdevitt2/RunAwayPINNs.

\section{\label{sec:DPR}Decay of a Primary Runaway Electron Distribution}

As a specific application of the above framework, we will treat the decay of an initial primary RE distribution. Here, an initial population of REs will be initialized with a given momentum space distribution and the adjoint-deep learning framework described in Secs. \ref{sec:AFR} and \ref{sec:PINN} will be used to project the RE density forward in time. We will begin by describing representative solutions to the adjoint of the relativistic Fokker-Planck equation (Sec. \ref{sec:TER}) and subsequently describe the decay of a primary RE population for a range of initial conditions (Sec. \ref{sec:REDR}). Finally, the predictions of the adjoint-deep learning framework will be compared with a traditional RE solver in Sec. \ref{sec:VAD}.

\subsection{\label{sec:TER}Temporal Evolution of the Runaway Probability Function}

\begin{figure}
\begin{centering}
\subfigure[]{\includegraphics[scale=0.5]{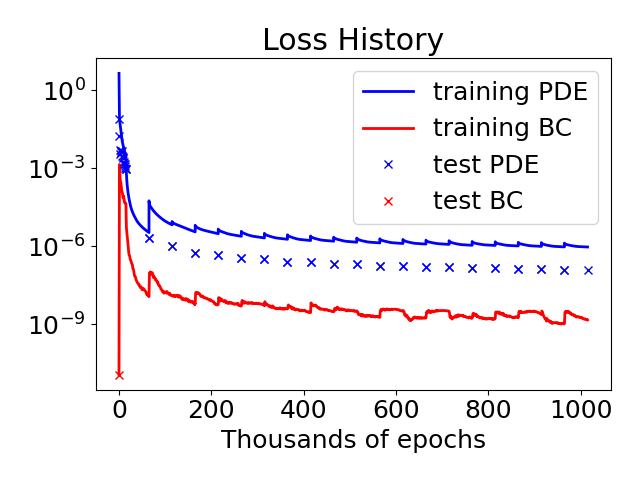}}
\subfigure[]{\includegraphics[scale=0.5]{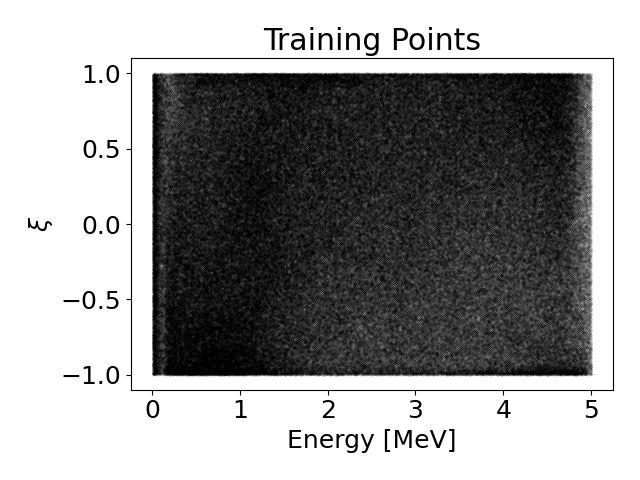}}
\par\end{centering}
\caption{(a) Training and test loss history. The solid blue curve indicates the training loss of the PDE, the solid red curve indicates the training loss for the boundary condition, and the `x' markers indicate the test loss (the test and training losses are the same for the boundary term). 1,000,000 training and testing points are used. The training points initially obey a Hammersley sequence, with the test distribution being uniform random. The training points are resampled using the residual based adaptive distribution described in Ref. \cite{wu2023comprehensive} every 50,000 epochs of L-BFGS. (b) Projection of the final training point distribution onto the energy and pitch domain.}
\label{fig:TER1}
\end{figure}

\begin{figure}
\begin{centering}
\subfigure[]{\includegraphics[scale=0.33]{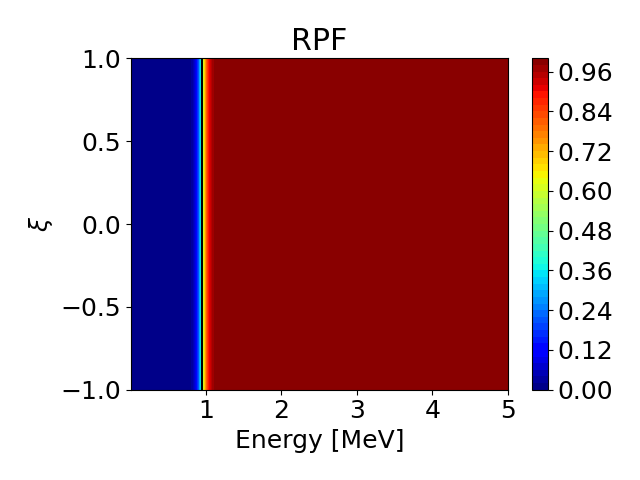}}
\subfigure[]{\includegraphics[scale=0.33]{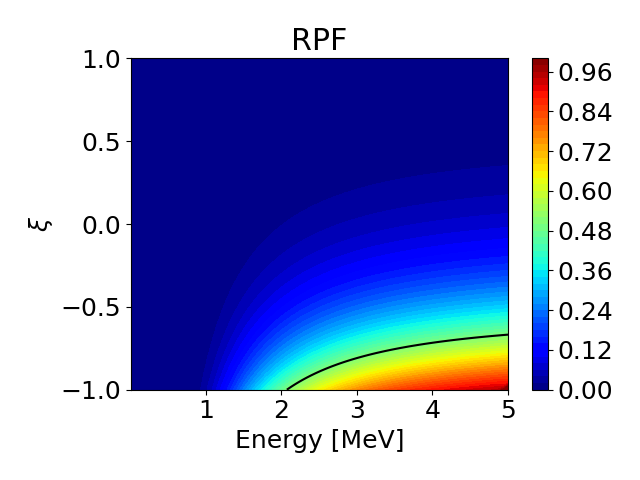}}
\subfigure[]{\includegraphics[scale=0.33]{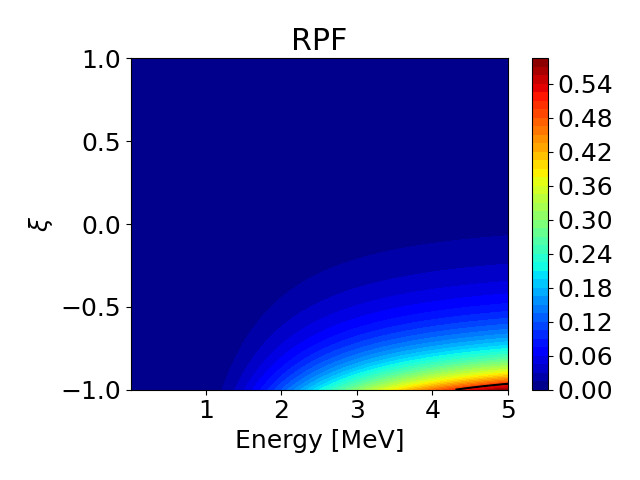}}
\subfigure[]{\includegraphics[scale=0.33]{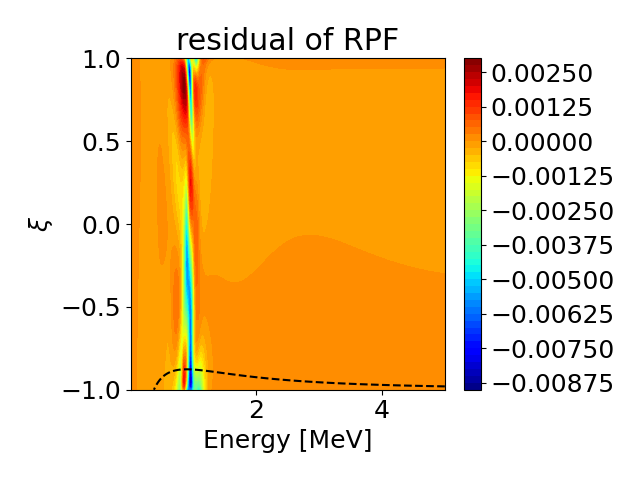}}
\subfigure[]{\includegraphics[scale=0.33]{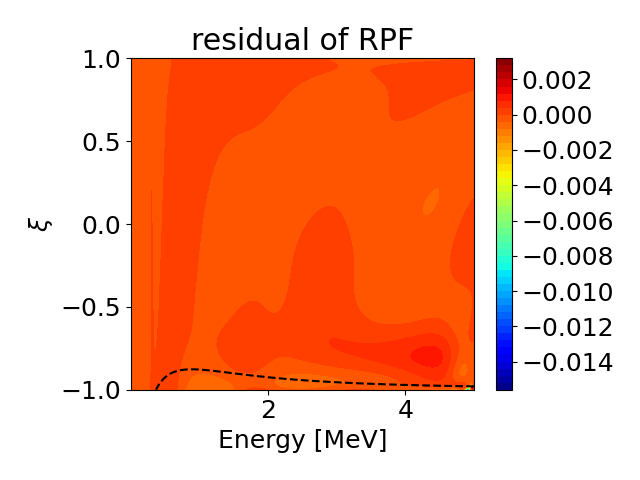}}
\subfigure[]{\includegraphics[scale=0.33]{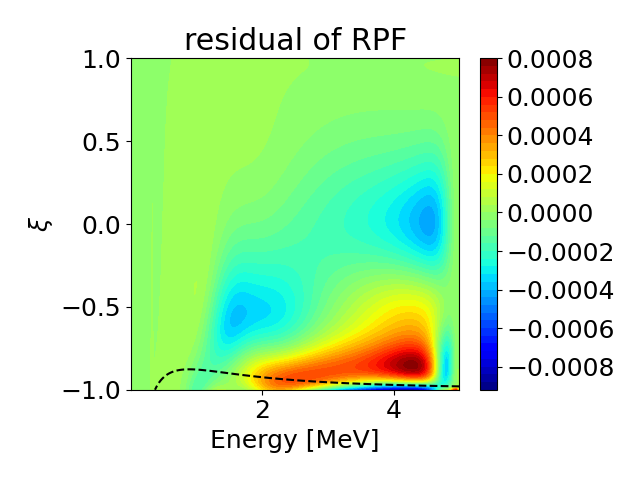}}
\par\end{centering}
\caption{Time slices of the RPF evolution (top row) along with the associated residual (bottom row) for $E_\Vert = 1.5$ and $p_{RE}=p_{max}/4$. The first column indicates the terminal condition $t=20$, the second column $t=10$ and the last column $t=0$. The other parameters are given by $Z_{eff}=2$ and $\alpha = 0.1$. The solid black contour in the top row indicates $P=0.5$ whereas the dashed black curve in the bottom row indicates the location of the $U_p = 0$ contour.}
\label{fig:TER2}
\end{figure}

\begin{figure}
\begin{centering}
\subfigure[]{\includegraphics[scale=0.33]{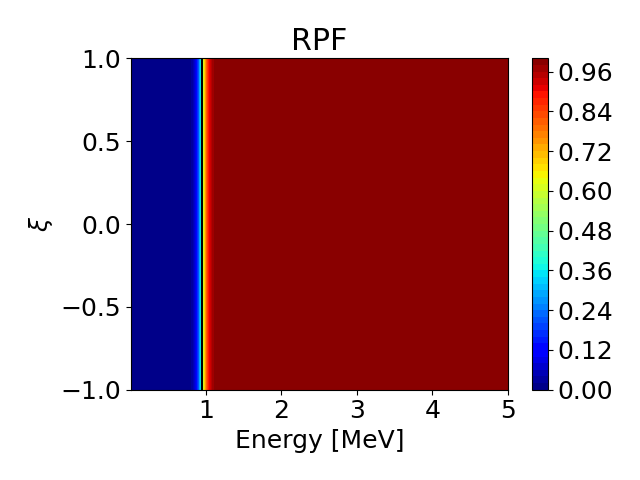}}
\subfigure[]{\includegraphics[scale=0.33]{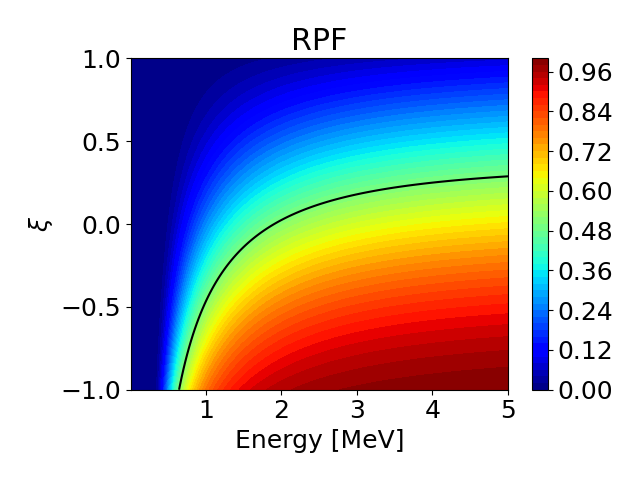}}
\subfigure[]{\includegraphics[scale=0.33]{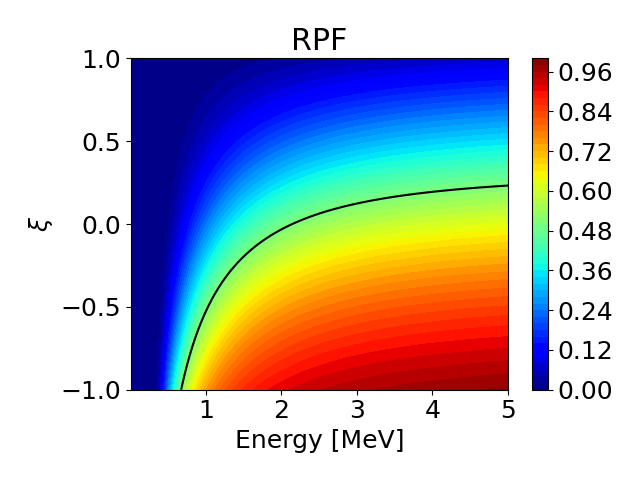}}
\subfigure[]{\includegraphics[scale=0.33]{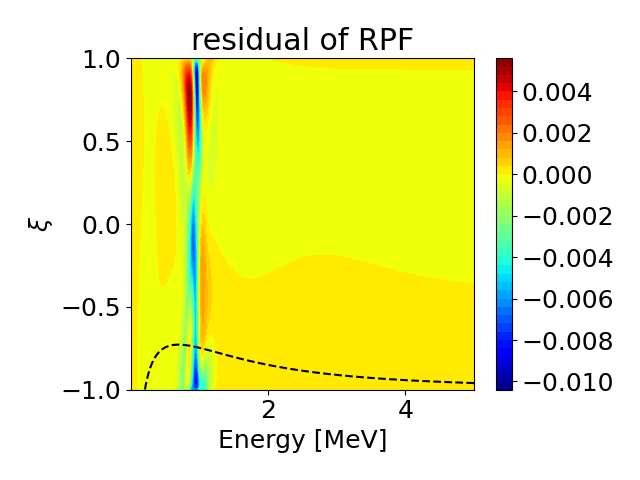}}
\subfigure[]{\includegraphics[scale=0.33]{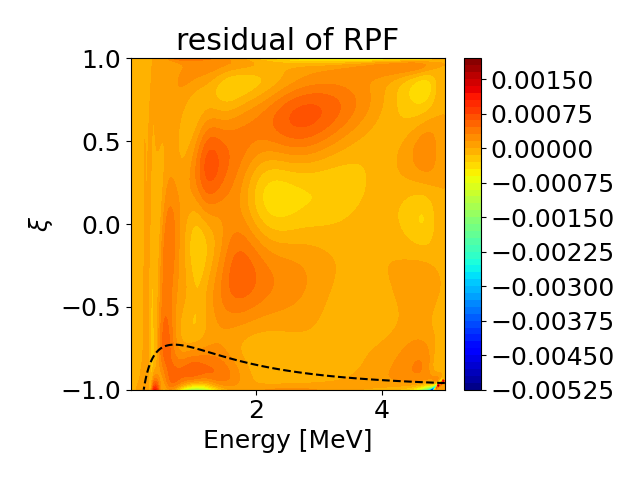}}
\subfigure[]{\includegraphics[scale=0.33]{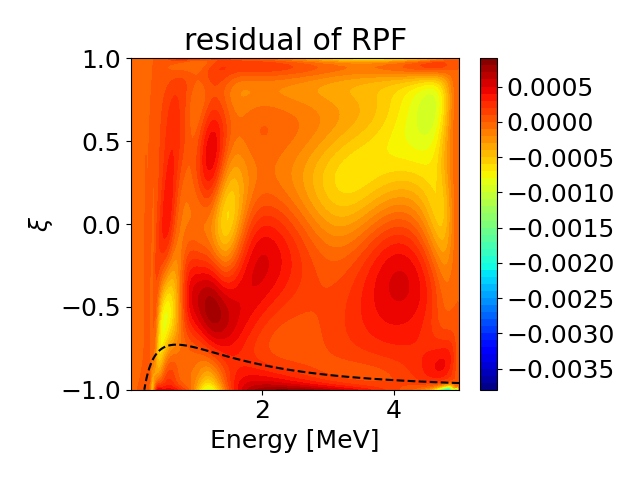}}
\par\end{centering}
\caption{Same as Fig. \ref{fig:TER2} but with $E_\Vert = 2$.}
\label{fig:TER3}
\end{figure}

Using the NN architecture and loss function described in Sec. \ref{sec:PRED} above, example solutions to the adjoint problem are shown in Figs. \ref{fig:TER2} and \ref{fig:TER3} with the test and training loss history shown in Fig. \ref{fig:TER1}. Here we have trained the PINN for electric fields in the range $E_\Vert \in \left(0, 3 \right)$, effective charges $Z_{eff} \in \left(1, 2 \right)$, and synchrotron radiation $\alpha \in \left( 0, 0.1\right)$. A much broader range of physics parameters will be considered in Ref. \cite{mcdevittpart22024}. We have also chosen a domain with a minimum energy of $10\;\text{keV}$, a maximum energy of $5\;\text{MeV}$, and $t_{final} =20$. The energy defining which electrons are considered to be REs is taken to be $p_{RE} = p_{max}/4$, which corresponds to an energy of roughly 1 MeV.
The loss history of the PINN is shown in Fig. \ref{fig:TER1}. Here, the first 15,000 epochs are performed using the ADAM optimizer, while L-BFGS is used thereafter. After periods of 50,000 epochs of training with L-BFGS, the training point distribution is resampled using the residual based adaptive distribution (RAD) method described in Ref. \cite{wu2023comprehensive}. Such a sampling routine redistributes roughly half of the training points using a uniform random distribution, with the remaining points selected with a probability proportional to the magnitude of the residual. In this way, the entire training domain is sampled, but with regions that have relatively large residuals receiving more training points [the final training point distribution is shown in Fig. \ref{fig:TER1}(b)]. This training point resampling results in spikes in the training loss every 50,000 epochs as evident from Fig. \ref{fig:TER3}(a). Despite these spikes in the training loss of the PDE, the test loss of the PDE monotonically decays to a value of $\approx 10^{-7}$, with the boundary loss term dropping to a lower value, suggesting that the PDE is well satisfied across the range of parameters under consideration.

Two example solutions are shown in Figs. \ref{fig:TER2} and \ref{fig:TER3} for electric fields below and above the avalanche threshold (the avalanche threshold is $E_\Vert \approx 1.8$ for these parameters). Considering the subthreshold case first ($E_\Vert = 1.5$, Fig. \ref{fig:TER2}), the $P=0.5$ contour (black contour in the top row of Fig. \ref{fig:TER2}) shifts to higher momentum and smaller pitch when moving from the $t= 20$ to $t=0$ time slice. Physically this implies that an electron requires an energy near 5 MeV and a pitch near $\xi=-1$ at $t=0$ to have a better than fifty percent chance of remaining in the RE region, or being accelerated through the high momentum boundary, by $t=t_{final}$. This is due to the combination of drag and synchrotron losses exceeding electric field acceleration throughout the majority of momentum space, with only a narrow channel where $U_p > 0$ (below the dashed black contour in the bottom row of Fig. \ref{fig:TER2}). Even inside this narrow channel, the presence of pitch-angle scattering will result in electrons initially inside the $U_p > 0$ region being scattered to a larger pitch-angle where $U_p < 0$, such that they are slowed down~\cite{Andersson:2001, Decker:2016, guo2017phase, mcdevitt2018relation} resulting in the RPF taking on a small value across nearly all of momentum space.

While the magnitude of the residual in Fig. \ref{fig:TER2} is small for the three time slices shown, suggesting an accurate solution, the final time slice [$t=0$, Fig. \ref{fig:TER3}(c)] suggests a limitation of the PINN approach employed. In particular, the high momentum boundary condition employed is $P \left( p=p_{max}\right) = 1$ when $U_p > 0$ [see Eq. (\ref{eq:RPF1c})]. This is not satisfied at $t=0$. Specifically, as evident from Fig. \ref{fig:TER2}(f) a narrow region near $\xi=-1$ is present where $U_p>0$. However, the RPF is not unity at the high momentum boundary in this narrow channel. The reason for this inconsistency is due to pitch-angle scattering quickly diffusing electrons out of the $U_p>0$ channel where they are subsequently slowed down. Thus, the magnitude of the RPF will quickly drop below unity for $p<p_{max}$ even when $U_p > 0$. Since the present PINN implements the high momentum boundary condition as a soft constraint, deviation from the prescribed boundary condition in this narrow boundary region only leads to a modest penalty in the boundary loss term. We also note that the region where the RPF deviates from the boundary condition ($p=p_{max}$, $\xi \approx -1$ and $t=t_{final}$) represents a corner in the training regime, it is thus likely that it is not directly sampled by the random distribution of training points. While this is indicative of a slight inconsistency in the PINN solution, we do not expect it to strongly impact predictions of the RE density since the PINN correctly captures the value of the RPF throughout the majority of the simulation domain, with only a small inconsistency present in the corner of the simulation domain when near marginality. This will be verified in Sec. \ref{sec:REDR}, where predictions of the PINN are compared with solutions from a traditional RE solver (see Fig. \ref{fig:VAD1}).

Considering now a case above threshold ($E_\Vert = 2$, Fig. \ref{fig:TER3}), the channel with $U_p>0$ is substantially broadened, facilitating the acceleration of electrons. In particular, the location of the $P=0.5$ contour near $\xi=-1$ shifts to lower momenta as time evolves, indicating more electrons will be accelerated by the electric field. Since the present case is only modestly above marginality, pitch-angle scattering introduces a substantial broadening of the transition region. In addition, the RPF reaches a near steady state condition after ten $\tau_c$, such that there is only a modest difference between the RPF at $t=10$ compared to $t=20$.

\subsection{\label{sec:REDR}Temporal Evolution of the Runaway Electron Density}

\subsubsection{Time Histories of RE Density for Different Initial Conditions}

\begin{figure}
\begin{centering}
\subfigure[]{\includegraphics[scale=0.33]{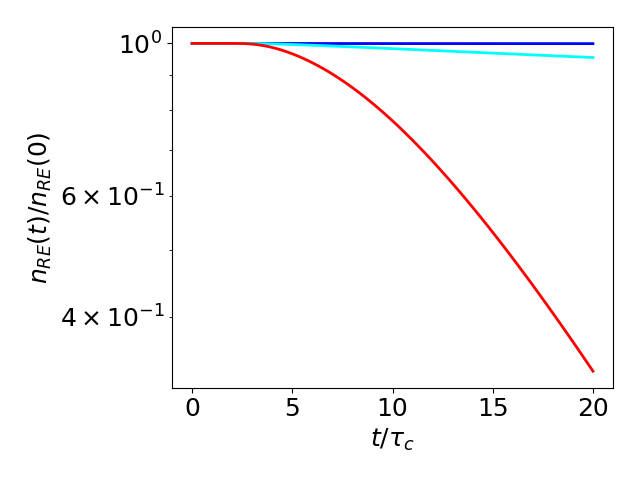}}
\subfigure[]{\includegraphics[scale=0.33]{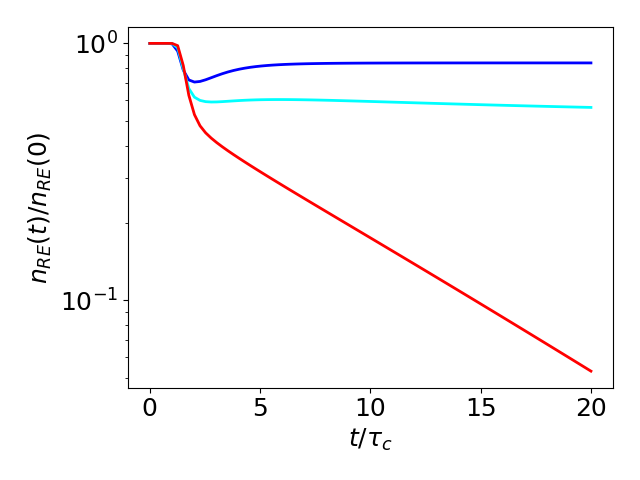}}
\subfigure[]{\includegraphics[scale=0.33]{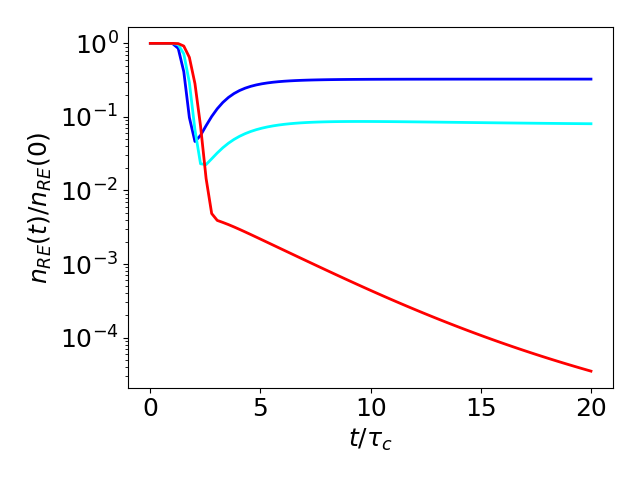}}
\subfigure[]{\includegraphics[scale=0.33]{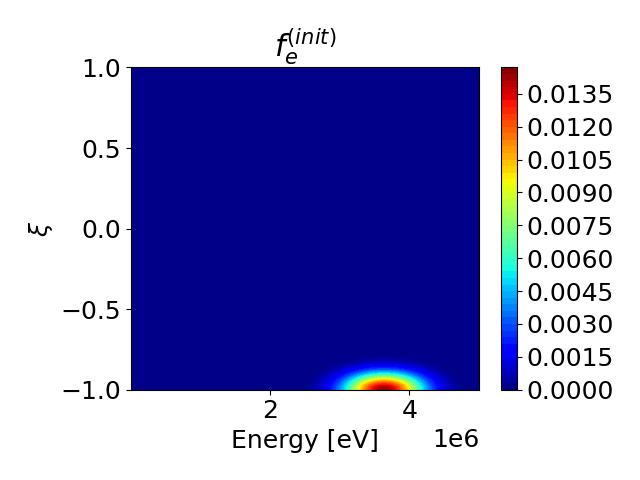}}
\subfigure[]{\includegraphics[scale=0.33]{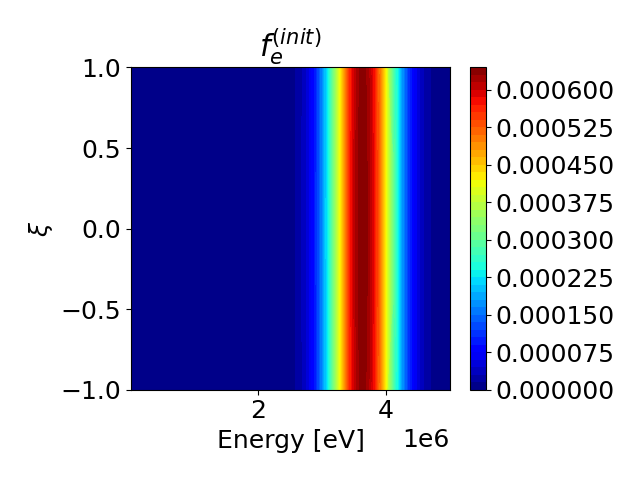}}
\subfigure[]{\includegraphics[scale=0.33]{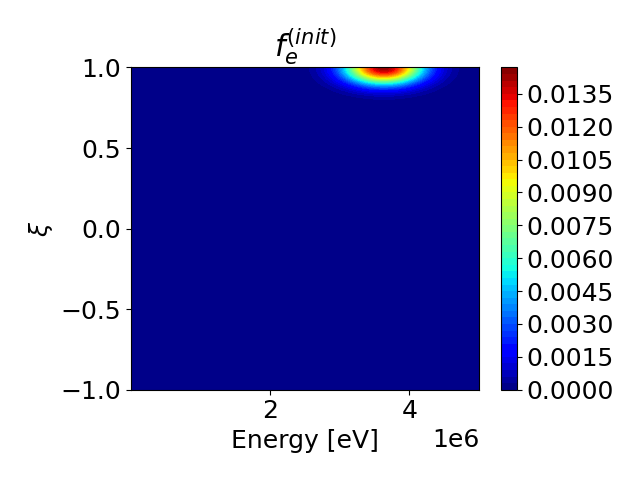}}
\par\end{centering}
\caption{Runaway electron density evolution for a range of electric fields and for the initial distribution defined by Eq. (\ref{eq:REDR2}) for $\left( \Delta \xi = 0.1,\; \xi_0 = -1\right)$ (panel a.), $\left( \Delta \xi = 10,\; \xi_0 = 0\right)$ (panel b.) and $\left( \Delta \xi = 0.1,\; \xi_0 = 1\right)$ (panel c.). Each case took $\left( \Delta p = 0.1 p_{max},\; p_0 = 0.75p_{max}\right)$. The electric fields selected correspond to $E_\Vert = 1.5$ (red curves), $E_\Vert = 2$ (cyan curves), and $E_\Vert = 2.5$ (blue curves). The other parameters are given by $Z_{eff}=2$ and $\alpha = 0.1$. The bottom row indicates the initial electron distributions.}
\label{fig:REDR1}
\end{figure}

In this section we will utilize the parametric solution to the time dependent RPF described in Sec. \ref{sec:TER} to evaluate the time history of the RE density $n_{RE}$ using Eq.  (\ref{eq:TDRP20}). 
Taking the initial RE population to be of the form:
\begin{equation}
f^{(init)}_e \left( p, \xi \right) \propto \exp \left[ -\frac{\left( p-p_0\right)^2}{\Delta p^2} -\frac{\left( \xi - \xi_0 \right)^2}{\Delta \xi^2} \right]
, \label{eq:REDR2}
\end{equation}
the time evolution of the electron density for three different initial electron distributions is shown in Fig. \ref{fig:REDR1}. Here for Fig. \ref{fig:REDR1}(a), the initial electron distribution has a pitch centered around $\xi=-1$ ($\xi_0=-1$ and $\Delta \xi=0.1$), which results in the RE number density remaining approximately constant for $E_\Vert > E_{av}$, and decaying otherwise (for these parameters $E_{av} \approx 1.8$). In contrast, for a nearly isotropic distribution [Fig. \ref{fig:REDR1}(b)], the number of REs initially decays for all electric field values, but then levels out later in time for cases with $E_\Vert > E_{av}$. Finally, for electrons initially with $\xi\approx 1$ [Fig. \ref{fig:REDR1}(c)], the number of REs decays rapidly for all electric field values, with the cases with strong electric fields decaying the fastest. This initial decay is due to the electric field slowing the electron population down together with drag and radiation. Later in time, cases with $E_\Vert > E_{av}$ again show a sharp increase in number and then level off, though the number of REs does not recover to its initial value. This latter property is due to electrons initially slowing to an energy below $p_{RE}$, but then being turned by a combination of the electric field and pitch-angle scattering, where they are subsequently reaccelerated along $\xi \approx -1$. Some electrons are lost to the low energy boundary, hence the magnitude of the RE population is reduced after the early transient phase.

\subsubsection{\label{sec:RED}Runaway Electron Decay Rate}

\begin{figure}
\begin{centering}
\subfigure[]{\includegraphics[scale=0.5]{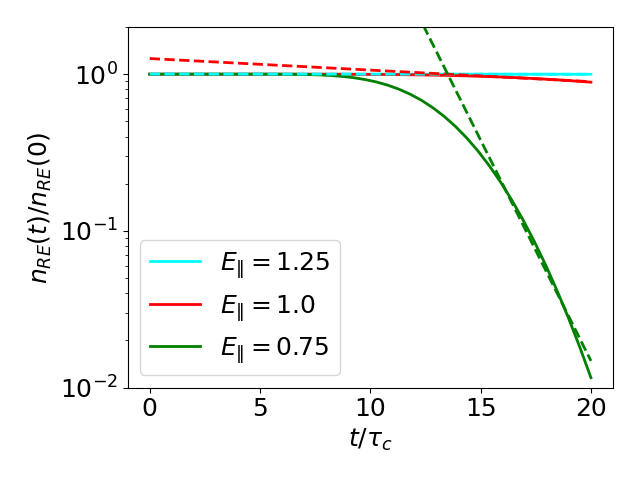}}
\subfigure[]{\includegraphics[scale=0.5]{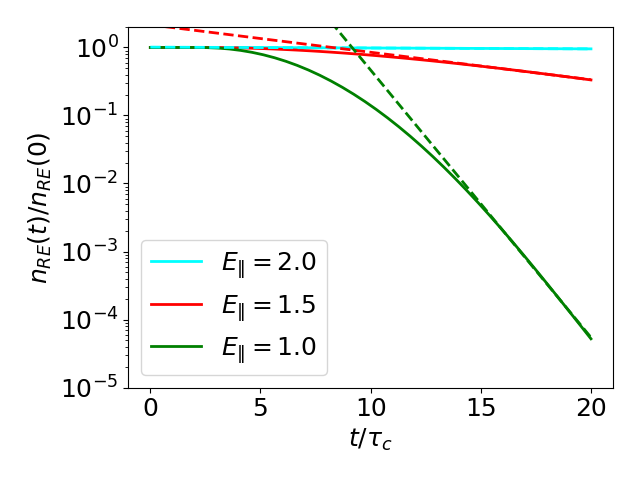}}
\par\end{centering}
\caption{The time evolution of the RE density for different electric fields. Panel (a) is for $Z_{eff}=1$ and $\alpha = 0$ where as panel (b) is for $Z_{eff}=2$ and $\alpha = 0.1$. The solid curves are predictions from the adjoint-deep learning framework, whereas the dashed curves are exponential fits. The last five $\tau_c$ were used in the exponential fits. The other parameters are given by $p_0 = 0.75p_{max}$, $\Delta p = 0.1 p_{max}$, $\xi_0 = -1$ and $\Delta \xi = 0.1$.
}
\label{fig:REDR1sub1}
\end{figure}

The time evolution of the RE density can be used to estimate the decay rate of a primary electron population when below threshold. Using a RE decay rate to characterize RE evolution comes with two significant caveats. The first is that while the RE density exhibits an exponential decay when slightly below threshold in the long time limit~\cite{mcdevitt2018relation}, its temporal evolution cannot be described by an exponential when the system is well below threshold.
As a concrete example, for an electric field $E_\Vert < 1$, the entire population of electrons will slow down due to collisional drag exceeding electric field acceleration at all energies and pitches. This will result in 
the primary population of electrons slowing down to the bulk in a finite amount of time, a process that cannot be described by a decaying exponential. This physics is illustrated in Fig. \ref{fig:REDR1sub1}(a), where the temporal evolution of the RE density for electric fields in the vicinity of $E_\Vert = 1$ for $Z_{eff}=1$ and $\alpha =0$ is shown. It is evident that for $E_\Vert < 1$ (green curve) an exponential curve cannot be fit to the RE density, as evidenced by the poor fit between the dashed and solid green curves. Furthermore, when at threshold ($E_\Vert = 1$, red curves) the exponential fit curve only approximates the latter portion of the time history of the RE density. Hence, a second caveat is that it often takes a substantial amount of time, tens to hundreds of $\tau_c$~\cite{mcdevitt2018relation}, for the REs to settle into a distribution with a clear exponential decay rate. Further noting that for a plasma with $n_e = 10^{20}\;\text{m}^{-3}$ and a Coulomb logarithm of $\ln \Lambda = 15$, the relativistic collision time is given by $\tau_c \approx 22\;\text{ms}$, implying that the RE density often will not achieve an exponential decay for tens to hundreds of milliseconds even for $E_\Vert \gtrsim 1$. Considering a case with $Z_{eff}=2$ and $\alpha = 0.1$ [Fig. \ref{fig:REDR1sub1}(b)], and thus a larger avalanche threshold of $E_{av} \approx 1.8$, a good exponential fit is not present for $E_\Vert = 1$ over twenty $\tau_c$, with only a marginally better fit present for $E_\Vert = 1.5$.

\begin{figure}
\begin{centering}
\subfigure[]{\includegraphics[scale=0.5]{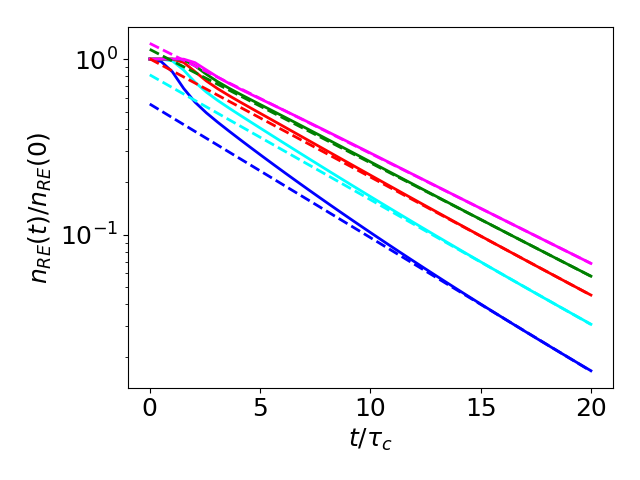}}
\subfigure[]{\includegraphics[scale=0.5]{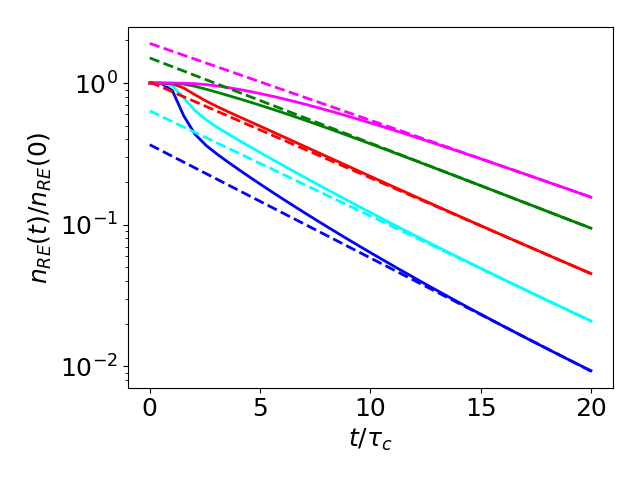}}
\par\end{centering}
\caption{The time evolution of the RE density for initial momentum space distributions centered about different momentum [panel (a)] and pitch [panel (b)]. For panel (a), the blue curve is for $p_0 = 0.4p_{max}$, the cyan curve is for $p_0 = 0.5p_{max}$, the red curve is for $p_0 = 0.6p_{max}$, the green curve is for $p_0 = 0.7p_{max}$ and the magenta curve is for $p_0 = 0.8p_{max}$. The pitch was taken to be $\xi_0=-0.5$ for all cases. For panel (b), the blue curve is for $\xi_0=-0.2$, the cyan curve is for $\xi_0=-0.4$, the red curve is for $\xi_0=-0.6$, the green curve is for $\xi_0=-0.8$ and the magenta curve is for $\xi_0=-1$. The momentum was taken to be $p_0=0.5p_{max}$ for all cases. The other parameters are given by $E_\Vert = 1.5$, $Z_{eff}=2$ and $\alpha = 0.1$, with the width of the momentum and pitch distributions given by $\Delta p = 0.1 p_{max}$ and $\Delta \xi = 0.1$, respectively.
}
\label{fig:REDR2}
\end{figure}

The time required for the RE distribution to settle into a well defined exponential decay depends sensitively on the initial momentum space distribution.
Specifically, while the inferred RE decay rate should be independent of the initial momentum space distribution, assuming the initial electrons are at sufficiently high energy, there will inevitably be a transient period while the RE momentum space distribution settles into its saturated form. To assess this uncertainty, a scan of the initial momentum and pitch of the primary electron distributions is shown in Figs. \ref{fig:REDR2}(a) and \ref{fig:REDR2}(b), respectively. Here it is evident that the early time evolution of the RE population varies substantially depending on both the initial momentum and pitch of the primary distribution, with resulting fitted exponentially implying a similar, though not identical, decay rate. The modest deviation in inferred decay rates is due to the finite time of the simulations, such that for initial conditions far from those that characterize the decaying primary distribution, twenty $\tau_c$ is not sufficient for the REs to settle into their saturated decay rate. A strength of the present approach is that by predicting the specific time history of the RE density, rather than only the decay rate, the accuracy of the exponential fit can be readily verified, and the specific time history can be used if required.

\begin{figure}
\begin{centering}
\subfigure[]{\includegraphics[scale=0.5]{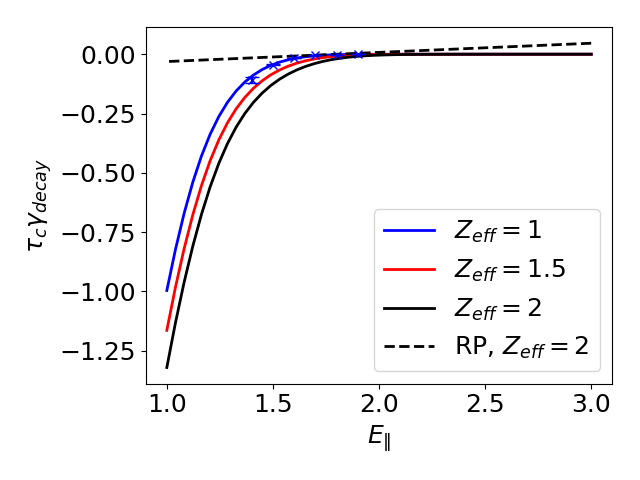}}
\subfigure[]{\includegraphics[scale=0.5]{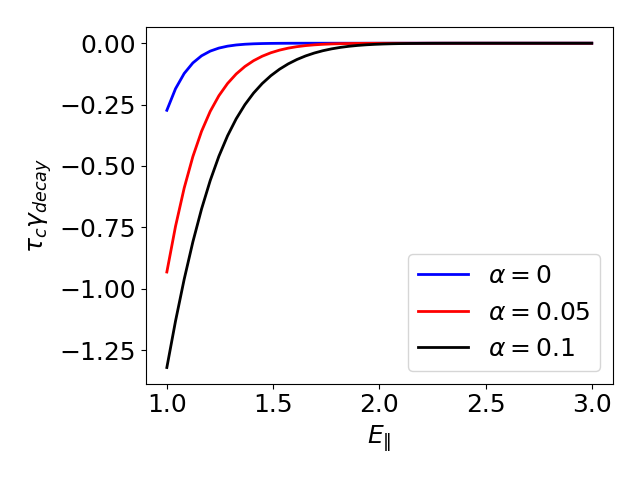}}
\subfigure[]{\includegraphics[scale=0.5]{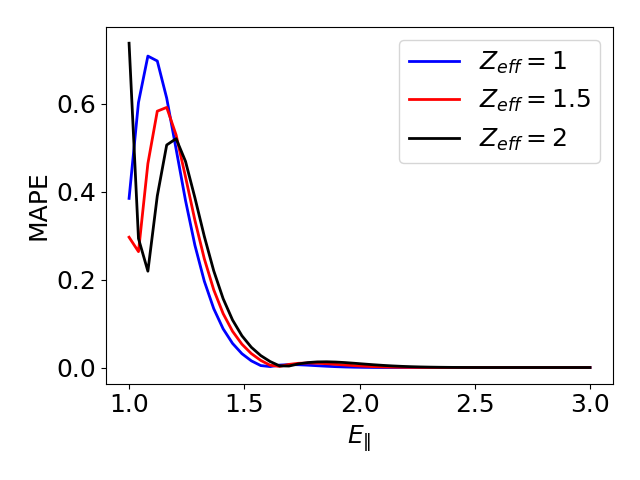}}
\subfigure[]{\includegraphics[scale=0.5]{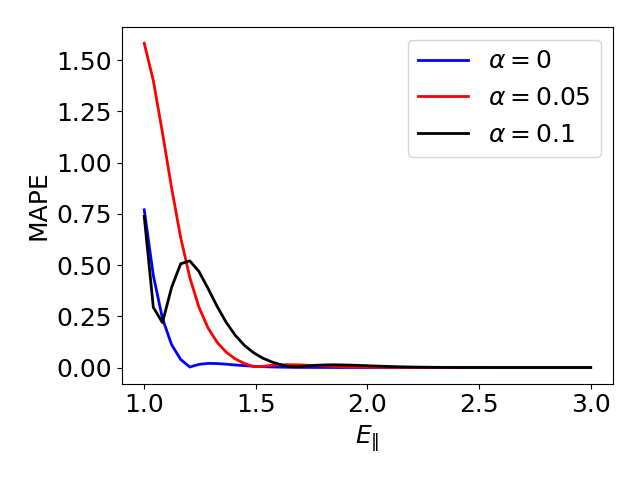}}
\par\end{centering}
\caption{Runaway electron decay rate versus electric field $E_\Vert$ for different effective charges $Z_{eff}$ [panel (a)] and synchrotron radiation $\alpha$ [panel (b)]. Panel (a) took $\alpha = 0.1$, whereas panel (b) took $Z_{eff} = 2$. The solid curves indicate predictions from the PINN, whereas the blue `x' markers are benchmark values taken from Ref. \cite{mcdevitt2018relation}. Panels (c) and (d) indicate the MAPE for the exponential functions used to estimate the RE decay rates in the first row. The initial momentum space distribution assumed $p_0 = 0.5p_{max}$, $\Delta p = 0.1 p_{max}$, $\xi_0 = -1$ and $\Delta \xi = 0.1$, respectively. The black dashed curve in panel (a) is Eq. (\ref{eq:REDR1}) for $\ln \Lambda = 10$, $Z_{eff}=2$, and $E_{av}=1.8$, which is consistent with a value of $\alpha=0.1$.}
\label{fig:REDR3}
\end{figure}

While an exponential decay rate is an imperfect measure of the RE density evolution, it does provide a convenient means of estimating the decay rate of a RE beam when below threshold.
A plot of the estimated decay rate of the primary electron distribution for a range of electric fields, effective charges and strength of synchrotron radiation is shown in Fig. \ref{fig:REDR3}. Here, time histories of the RE density were generated for twenty $\tau_c$, where the last five $\tau_c$ were used to estimate an effective decay rate. The decay rate nearly vanishes for electric fields greater than $E_\Vert \approx 2$, due to the system being above the avalanche threshold for those cases. For smaller electric fields, the magnitude of the decay rate increases rapidly, illustrating a strong nonlinear dependence on electric field strength. As a means of identifying the validity of the exponential fits, we have also plotted the Mean Absolute Percentage Error (MAPE) defined by:
\[
MAPE = \frac{100}{N} \sum^N_{i=1} \left| \frac{n_{RE,i} - n_{fit,i}}{n_{RE,i}}\right|
,
\]
where $n_{RE,i}$ are the values of RE density predicted by the adjoint-deep learning framework, $n_{fit,i}$ indicates the exponential fit, and $N$ is the number of points used to fit the exponential decay rate. Values of the MAPE are shown in the second row of Fig. \ref{fig:REDR3} as a function of electric field strength. It is evident that for electric fields substantially above the avalanche threshold, the MAPE reaches an exceptionally small value. This is due to the number density of REs being nearly constant in this region which is well fit by the negligibly small decay rates inferred. However, as the electric field is reduced below threshold, the magnitude of the associated MAPE increases rapidly. While the percent error never reaches a large percentage (the maximum MAPE is less than $2\%$) this is due to the relatively short time increment used to fit data such that deviations between the data and fit are never too large. For reference, a case with an MAPE of $\approx 0.75$ and parameters $E_\Vert = 1,\; Z_{eff}=2,\; \alpha = 0.1$, is shown by the green curve in Fig. \ref{fig:REDR1sub1}(b), where a rather poor fit is evident.
We thus anticipate that the inferred exponential decay rates will be most accurate when only weakly below threshold. For cases with significant MAPE values, large uncertainties are expected due to a decaying exponential providing a poor fit to the RE density over the twenty $\tau_c$ timescale employed. 

\begin{figure}
\begin{centering}
\includegraphics[scale=0.5]{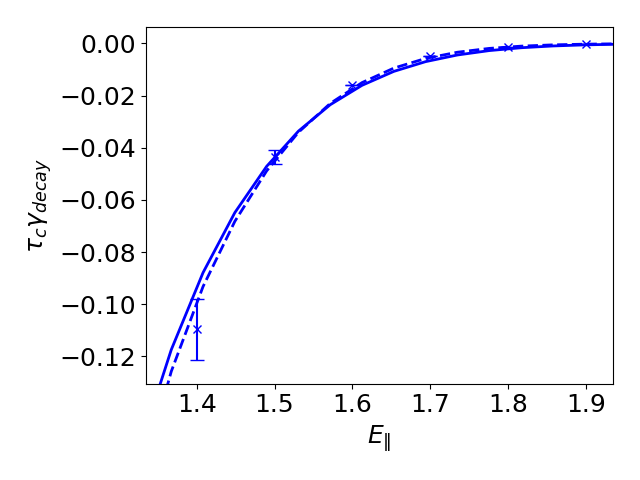}
\par\end{centering}
\caption{Comparison of decay rates predicted by a model with $t_{final}=20$ (solid blue curve) and $t_{final}=40$ (dashed blue curve) with benchmark results from Ref. \cite{mcdevitt2018relation} (`x' markers). The parameters were taken to be $Z_{eff}=1$, $\alpha = 0.1$ and $p_0 = 0.5p_{max}$, $\Delta p = 0.1 p_{max}$, $\xi_0 = -1$ and $\Delta \xi = 0.1$.}
\label{fig:REDR4}
\end{figure}

Insight into the accuracy of the exponential decay rate near threshold can be gained by comparing with RE decay rates inferred in Ref. \cite{mcdevitt2018relation}. Here, the RE solver RAMc~\cite{mcdevitt2019runaway} was used to evolve the primary RE distribution for $250\tau_c$, leading to well converged exponential decay rates. A comparison between the PINN and results from Ref. \cite{mcdevitt2018relation} are shown by the blue `x' markers in Fig. \ref{fig:REDR3}(a) and \ref{fig:REDR4}. The solid blue curve in both figures indicates results from a PINN trained with $t_{final}=20\tau_c$, whereas the dashed blue curve in Fig. \ref{fig:REDR4} being an analogous PINN, but trained over a time period of $t_{final}=40\tau_c$. Twice as many training points were used in this latter case (2,000,000) resulting in an increase in offline training time. It is evident that both results are in good agreement with the benchmark results, though the PINN with the larger $t_{final}$ exhibits modestly improved agreement.

The inferred RE decay rates differ substantially from commonly employed analytic expressions. Taking the well known approximate analytic expression for RE growth and decay in the completely screened limit defined by~\cite{Rosenbluth:1997}:
\begin{equation}
\tau_c \gamma_{av} = \frac{1}{\ln \Lambda} \sqrt{\frac{\pi}{3\left(Z_{eff}+5\right)}} \left( E_\Vert - E_{av}\right)
, \label{eq:REDR1}
\end{equation}
as an example, this expression is shown by the dashed black curve in Fig. \ref{fig:REDR3}(a). Here, we have modified Eq. (\ref{eq:REDR1}) to include a non-unity avalanche threshold, to account for effects of synchrotron radiation. The dashed curve was plotted for a Coulomb logarithm of $\ln \Lambda = 10$, $Z_{eff}=2$, and we chose $E_{av}=1.8$, consistent with $\alpha =0.1$. 
From the dashed black curve Fig. \ref{fig:REDR3} it is evident that this expression gives a decay rate that drastically underestimates the magnitude of RE decay compared to the prediction of the adjoint-deep learning framework, even when the avalanche threshold is adjusted. We note that while the precise magnitude of the strongly decaying cases is subject to substantial uncertainties, as described above, the weak exponential decay rates predicted by Eq. (\ref{eq:REDR1}) deviate strongly from the benchmark RE decay rates (`x' markers). Furthermore, while the present adjoint-deep learning framework does not include large-angle collisions, the companion paper Ref. \cite{mcdevittpart22024} demonstrates that the decay rate including large-angle collisions is well approximated by the decay rate of the primary population. We thus anticipate simple analytic expressions such as Eq. (\ref{eq:REDR1}) to substantially underestimate the decay of REs when below threshold.


\subsection{\label{sec:VAD}Verification of Adjoint-Deep Learning Framework}

\begin{figure}
\begin{centering}
\subfigure[]{\includegraphics[scale=0.33]{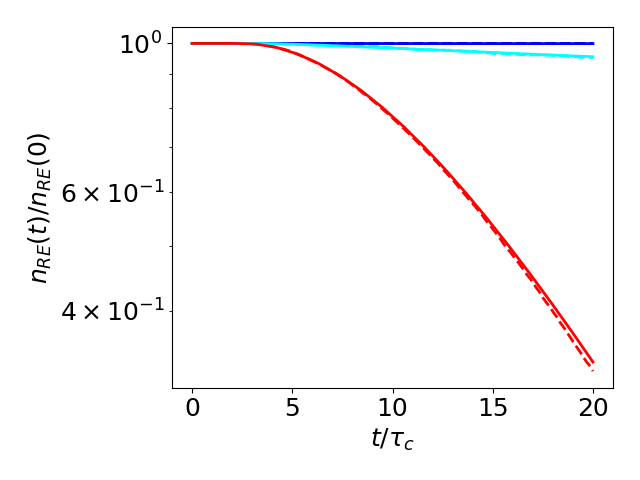}}
\subfigure[]{\includegraphics[scale=0.33]{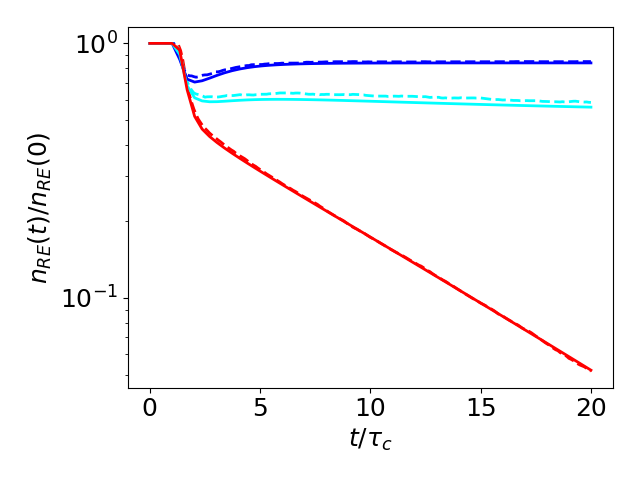}}
\subfigure[]{\includegraphics[scale=0.33]{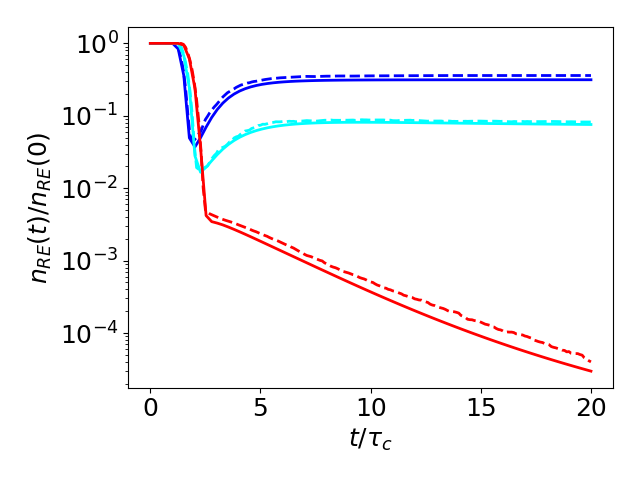}}
\subfigure[]{\includegraphics[scale=0.33]{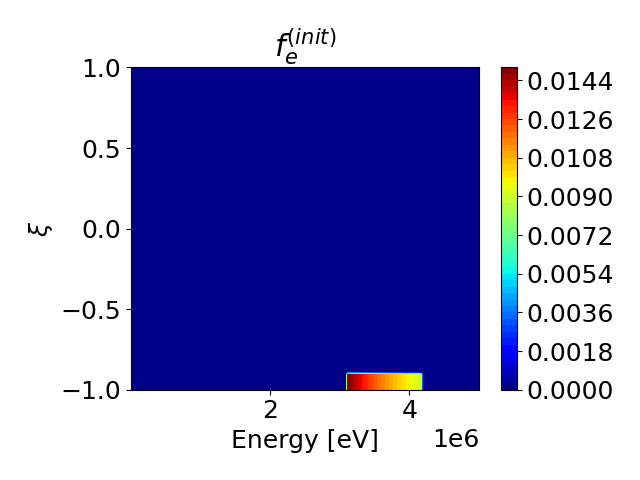}}
\subfigure[]{\includegraphics[scale=0.33]{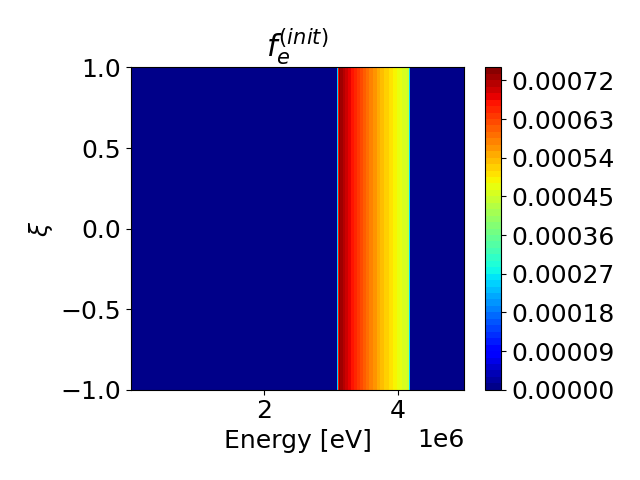}}
\subfigure[]{\includegraphics[scale=0.33]{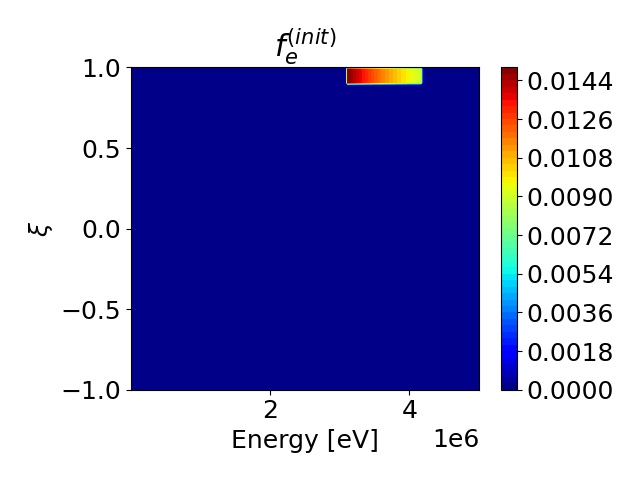}}
\par\end{centering}
\caption{The top row shows a comparison of the PINN's prediction of the RE density evolution (solid curves) with the Monte Carlo code RAMc (dashed curves). The bottom row indicates the initial primary distribution. The electric fields selected correspond to $E_\Vert = 1.5$ (red curves), $E_\Vert = 2$ (cyan curves), and $E_\Vert = 2.5$ (blue curves). The other parameters are given by $Z_{eff}=2$ and $\alpha = 0.1$. The bottom row indicates the initial electron distributions.}
\label{fig:VAD1}
\end{figure}

As a means of verifying predictions of the adjoint-deep learning framework we will compute several time histories of the RE density using the Monte Carlo code RAMc~\cite{mcdevitt2019runaway} for several different initial momentum space distributions and physical parameters. While a comparison of inferred RE decay rates could readily be performed, due to the ambiguities that often emerge when inferring the RE decay rate from the RE time history (as described in Sec. \ref{sec:RED}), a more stringent test will be to compare detailed time histories of the RE density. A comparison of several temporal histories of RE density with the Monte Carlo code RAMc~\cite{mcdevitt2019runaway} is illustrated in Fig. \ref{fig:VAD1}. Here, we have used the initial momentum space distributions illustrated by the second row of Fig. \ref{fig:VAD1}, and the values of $\alpha$ and $Z_{eff}$ are the same as the RPFs shown in Fig. \ref{fig:TER2} above. The form of these momentum space distributions were chosen to be consistent with the particle initialization scheme used in RAMc. Good agreement is evident for the three electric fields and three initial conditions considered. For the practically important case of an initial RE distribution approximately aligned with the magnetic field ($\xi = - 1$) the adjoint-deep learning predictions are in excellent agreement with the RAMc code. For cases with electrons further from $\xi = -1$, progressively larger deviations are present, though good overall agreement is evident with the largest deviation appearing for the lowest electric field with electrons initialized with $\xi = 1$ [see Fig. \ref{fig:VAD1}(c) (red curves)]. The origin of this deviation is due to the limited accuracy of the PINN's prediction of the RPF in regions where the RPF is nearly vanishing. In particular, the small number of REs evident in the solid red curve in Fig. \ref{fig:VAD1}(c) results from integrating
the product of the RPF shown in Fig. \ref{fig:TER2}(c) with the initial momentum space distribution shown in Fig. \ref{fig:VAD1}(f). Since the RPF is nearly zero in the region where the initial electron distribution is finite (i.e. at $\xi\approx1$ and $E\approx 3.5\;\text{MeV}$), 
this thus requires the RPF to be computed very accurately to resolve this near zero value. As evident from Fig. \ref{fig:VAD1}(c), the PINN is able to accurately track the initial density decay, but exhibits a quantitative discrepancy once only a trace RE population remains.


\section{\label{sec:C}Conclusion}

An adjoint-deep learning framework for evaluating the time evolution of the density moment of the RE distribution was derived. By solving the adjoint to the relativistic Fokker-Planck equation this allows the RE density to be evolved from an arbitrary initial momentum space distribution for a given set of parameters. By utilizing a PINN to identify the parametric solution to the adjoint problem, this further generalizes the framework to treat a range of plasma conditions. While this framework only evolves the RE density moment, by solving the adjoint to the relativistic Fokker-Planck equation, this enables a fully kinetic treatment of relativistic electron dynamics. As an initial application, the evolution of a primary electron population was evolved from three distinct initial pitch distributions, for three different values of the electric field, yielding good agreement with a traditional RE solver.
While the offline training time was substantially longer than a traditional relativistic Fokker-Planck solver, the online prediction time is substantially less, typically only requiring tens to hundreds of milliseconds. Furthermore, the decay rate of primary electrons was estimated as a function of electric field strength, effective charge, and the strength of synchrotron radiation. The estimated decay rate was found to vary nonlinearly with the electric field strength, in sharp contrast to commonly employed analytic theories~\cite{Rosenbluth:1997} that yield a linear dependence on electric field strength. This stiff electric field dependence implies that the electric field will need to remain near threshold during the RE plateau~\cite{Breizman:2014, mcdevitt2023runaway}, to sustain the RE population. While substantial ambiguities in the use of a RE decay rate for characterizing RE evolution when below threshold were identified, by predicting specific time histories of the RE density, the present approach allows for such limitations to be quantified. In particular, aside from predicting the RE decay rate, the prediction of the RE time history enables the quality of the exponential fit to be readily evaluated, where the residual of the PDE provides further insight into the accuracy of the solution.



Despite these encouraging initial results, several limitations of the adjoint-deep learning framework are apparent. One limitation in the present approach is the use of collisional coefficients that are only valid in the completely screened limit. Weakly ionized impurities are typically present in the plasma during a tokamak disruption and modify the collisional processes between electrons~\cite{Hesslow:2017}. While accounting for these partial screening effects will require modifying the PINN, we anticipate a straightforward extension of the existing PINN and is ongoing work. A second limitation, evident from Fig. \ref{fig:REDR2}(c), is the modest accuracy that PINNs are able to achieve when solving a PDE in comparison to traditional Fokker-Planck solvers~\cite{Harvey:2000, Nilsson:2015, guo2017phase, hoppe2021dream}. In particular, for the case shown in Fig. \ref{fig:VAD1}(c) the initial RE population has decayed by several orders of magnitude, such that the predicted number of REs is determined by a region of momentum space where the RPF is nearly zero. Since such regions where $P \approx 0$ yield small contributions to the loss function defined by Eq. (\ref{eq:PCDL2}), the PINN will not be able to accurately resolve the RPF in these regions without special care. 
A third limitation is that the range of parameters that the PINN was trained across in this paper is fairly narrow [$E_\Vert \in \left( 0,3 \right)$, $Z_{eff} \in \left( 1,2 \right)$, $\alpha \in \left( 0, 0.1\right)$], and thus will not be able to describe a broad range of tokamak conditions. This narrow training regime was selected to provide a conceptually clear treatment of the adjoint-deep learning framework, while avoiding the technical complexities that emerge when training across a broad range of parameters. This limitation is addressed in the companion paper Ref. \cite{mcdevittpart22024}, where an adaptive energy and time domain are introduced that enable a far broader parameter regime to be considered while using a comparable number of training points. Finally, an accurate description of RE evolution when above threshold requires the inclusion of large-angle collisions in order to describe the exponential growth of the RE population. This limitation is addressed in the companion paper Ref. \cite{mcdevittpart22024}, where a Rosenbluth-Putvinski secondary source term is used to approximate large-angle collisions.


\begin{acknowledgements}

  This work was supported by the Department of Energy, Office of Fusion Energy Sciences
at the University of Florida
  under awards DE-SC0024649 and DE-SC0024634, and at Los Alamos National Laboratory (LANL) under contract No. 89233218CNA000001. The authors acknowledge the University of Florida Research Computing for providing computational resources that have contributed to the research results reported in this publication. This research used resources of the National Energy Research Scientific Computing Center (NERSC), a Department of Energy Office of Science User Facility using NERSC award FES-ERCAP0028155.

\end{acknowledgements}

%
%
%
%


\end{document}